\documentclass[10pt,aps,prl,twocolumn,superscriptaddress,floatfix,nofootinbib]{revtex4-1}
\usepackage{times}
\usepackage[T1]{fontenc}

\usepackage{multirow}

\usepackage[version=4]{mhchem}

\usepackage{ulem}

\usepackage{amsmath,amssymb,mathrsfs}
\usepackage{bm}
\usepackage{ascmac}
\usepackage{geometry}
\usepackage{graphicx}

\def\0\\{\nonumber\\}

\makeatletter
\newcommand\footnoteref[1]{\protected@xdef\@thefnmark{\ref{#1}}\@footnotemark}
\makeatother

\begin{document}

\title{
Possible Existence of Extremely Neutron-Rich Superheavy Nuclei in Neutron Star Crusts\\
Under a Superstrong Magnetic Field
}

\author{Kazuyuki Sekizawa}
\email[]{sekizawa@phys.titech.ac.jp}
\affiliation{Department of Physics, School of Science, Tokyo Institute of Technology, Tokyo 152-8551, Japan}
\affiliation{Nuclear Physics Division, Center for Computational Sciences, University of Tsukuba, Ibaraki 305-8577, Japan}
\affiliation{RIKEN Nishina Center, Saitama 351-0198, Japan}

\author{Kentaro Kaba}
\affiliation{Department of Physics, School of Science, Tokyo Institute of Technology, Tokyo 152-8551, Japan}

\date{February 16, 2023}

\begin{abstract}
We investigate outer crust compositions for a wide range of magnetic field strengths, up to
$B$\,$\simeq$\,4\,$\times$\,10$^{18}$\,G, employing the latest experimental nuclear masses supplemented
with various mass models. The essential effects of the magnetic field are twoholds: 1) Enhancement of
electron fraction, which is connected to that of protons via the charge neutrality condition, due to
the Landau-Rabi quantization of electron motion perpendicular to the field, namely, neutron-richness is
supressed for a given pressure. As a result, 2) nuclei can exist at higher pressures without dripping
out neutrons. By exploring optimal outer-crust compositions from all possible nuclei predicted by
theoretical models, we find that neutron-rich heavy nuclei with neutron magic numbers 50, 82, 126,
as well as 184, with various proton numbers emerge for $B$\,$\gtrsim$\,10$^{17}$\,G. Moreover, we show
that superheavy nuclei with proton numbers $Z$\,$\ge$\,104, including unknown elements such as $Z$\,$=$\,119,
120, 122, and/or 124, depending on mass models, may emerge as an equilibrium composition at bottom layers
of the outer crust for $B$\,$\gtrsim$\,10$^{18}$\,G, which are extremely neutron-rich ($N$\,$\approx$\,260--287,
\textit{i.e.}, $N/Z$\,$\approx$\,2.2--2.4). We point out that those extremely neutron-rich superheavy nuclei locate
around the next neutron magic number $N$\,$=$\,258 after $N$\,$=$\,184, underlining importance of nuclear
structure calculations under such really exotic, extreme conditions. We demonstrate how the superstrong
magnetic field substantially alters crustal properties of neutron stars, which may have detectable consequences.
\end{abstract}


\maketitle

\begin{figure} [t]
\includegraphics[width=\columnwidth]{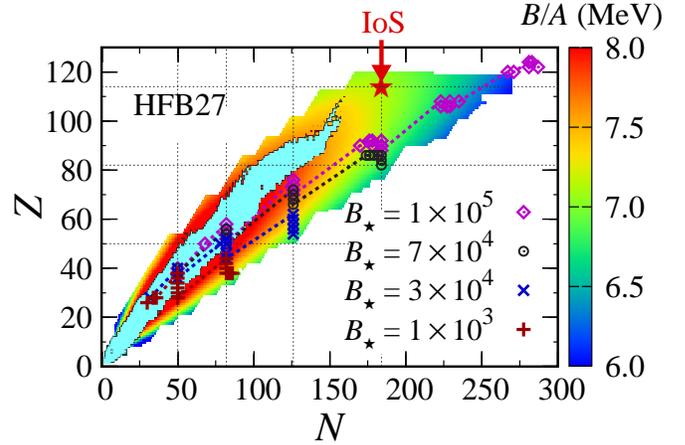}\vspace{-3mm}
\caption{
The main finding of this Letter. Colored area shows binding energy per nucleon
($B/A$ in MeV) predicted by the microscopic HFB-27 mass model \cite{HFB27}.
Shaded area indicates the region where experimental masses are available in AME2020
\cite{AME2020-I,AME2020-II}. On top of that, calculated equilibrium compositions
of the outer crust in strongly-magnetized neutron stars 
are plotted for $B_\star$\,$=$\,10$^{5}$ (diamonds), 7\,$\times$10$^4$ (circles),
3$\times$\,10$^4$ (crosses), and 10$^3$ (plus symbols), where $B_\star$\,$=$\,$B/B_\text{c}$
with $B_\text{c}$\,$\simeq$\,4.4\,$\times$\,10$^{13}$\,G. Dotted lines connect those
symbols to guide the eye. Vertical (horizontal) dotted lines indicate neutron
(proton) spherical magic numbers, $N$\,$=$\,50, 82, 126, and 184 ($Z$\,$=$\,50,
82, and 114). The \textit{island of stability} (IoS) located around $Z$\,$=$\,114
and $N$\,$=$\,184 is indicated by a star. Emergence of extremely $N$-rich nuclei
is clearly exhibited (see texts for details).
}\vspace{-3mm}
\label{Fig1}
\end{figure}

\textit{Introduction.---}What is the heaviest element that can, if it is very short lived,
exist in nature? This simple, yet profound question has urged us to synthesize unknown
superheavy nuclei (SHN) at terrestrial accelerator facilities all over the world
\cite{Hofmann(2000),Armbruster(2000),Hamilton(2013),Oganessian(2017),Giuliani(2019)}.
The synthesis of SHN is notoriously difficult due to its really tiny cross sections on
the order of picobarn to femtobarn. Nevertheless, owing to the continuous endeavor for
synthesizing SHN, the elements up to the atomic number $Z$\,$=$\,113, nihonium, and
$Z$\,$=$\,118, oganesson, have been synthesized so far via the so-called cold and hot
fusion reactions, respectively. Although the artificially synthesized SHN are short-lived,
their chemical as well as nuclear properties are of great interests, offering a unique opportunity
to challenge our theoretical understanding. The ever-ending quest for the next unknown elements
119, 120, and beyond, is the most challenging subject in nuclear physics today. In this Letter
we show that extremely neutron-rich ($N$-rich) SHN, including the yet unknown elements,
may naturally exist as an equilibrium composition of solid crusts of strongly-magnetized
neutron stars; see~Fig.~\ref{Fig1}.

Neutron stars are known to sustain a very strong magnetic field. The surface magnetic field
in the majority of pulsars is about $B$\,$\approx$\,10$^{11}$--10$^{13}$\,G, but some radio
pulsars exhibit larger magnetic fields on the order of 10$^{13}$--10$^{14}$\,G \cite{NeutronStars1}.
Further, from observations of soft $\gamma$-ray repeaters (SGRs) and anomalous x-ray pulsars
(AXPs), stronger surface magnetic fields up to 2.4\,$\times$\,10$^{15}$\,G have been inferred
\cite{Seiradakis(2004),Ng(2011),Mereghetti(2008),Olausen(2014),Tiengo(2013)}. According to
the virial theorem and magnetohydrodynamics simulations, the upper limit on the neutron-star
magnetic fields could be on the order of $B$\,$\approx$\,10$^{18}$\,G \cite{Potekhin(1996),
Potekhin(1999),Broderick(2000),Ventura(2001)}. Although it would be a rare situation, it is
worth exploring neutron star properties under such an extreme condition, prior to possible
observational finding. In this work, we thus focus on effects of a magnetic field on the
outer crust compositions up to the extreme condition, $B$\,$\gtrsim$\,10$^{18}$\,G, instead
of higher density regions where filling of higher Landau levels weakens the effects.

Effects of a magnetic field on outer crust compositions have been studied with various
models, such as Thomas-Fermi model \cite{Fushiki(1989)}, liquid drop model \cite{Lai(1991)},
Nilsson model \cite{Kondratyev(2000),Kondratyev(2001)}, covariant (relativistic) density functional
theory (CDFT) \cite{Arteaga(2011),Basilico(2015)}, Skyrme (non-relativistic) DFT \cite{Chamel(2012),
Chamel(2020)2}, effective relativistic mean-field model \cite{Parmar(2022)}. The common understanding
is that the most significant effect of a magnetic field originates from the Landau-Rabi quantization
of electron motion perpendicular to the field \cite{Rabi(1928),Landau(1930)}. In addition, nuclear
structure could be affected for $B$\,$\gtrsim$\,10$^{17}$\,G, where spin-dependent shifts of single-particle
energies become on the order of MeV \cite{Arteaga(2011),Basilico(2015),Stein(2016)_matter,Stein(2016)_C-O-Ne}.
In such a case, the size of shell gaps and even the ordering of energy levels may be altered and,
as a result, structural properties, such as deformation, radius, mass, could be affected. However,
the studies carried out so far are limited to a relatively light mass region, below or around
Fe up to around Xe \cite{Parmar(2022)} or so, probably because nuclides treated as candidates of
an equilibrium composition were somehow restricted.

In this Letter, we shall extend the investigation along the line with Ref.~\cite{Chamel(2012)},
adopting the latest experimental masses, AME2020 \cite{AME2020-I,AME2020-II}, supplemented with
various mass models (HFB-27 \cite{HFB27}, HFB-32 \cite{HFB30-32}, BML \cite{BML}, and KTUY \cite{KTUY}).
We carry out a comprehensive search for the equilibrium compositions from all nuclei in the entire
nuclear chart, \textit{i.e.} until the drip lines including SHN, for a wide range of magnetic
field strengths. We show that extremely $N$-rich SHN are allowed to exist as an equilibrium
composition of bottom layers of the outer crust of strongly-magnetized neutron stars. We mention
here that similar possibilities of emergence of SHN have been discussed in the context of supernova
explosions, where large electron fractions could be realized in an explosive situation (see,
\textit{e.g.}, Ref.~\cite{Iida(2020)}). We discuss significant changes of crustal properties
caused by the superstrong magnetic field, which may have detectable consequences.

\textit{Method.---}
The most significant effect of the magnetic field is caused by the Landau-Rabi quantization of
electron motion perpendicular to the field \cite{Rabi(1928),Landau(1930)}. The Landau energy level
of relativistic electrons is given by $E_\nu(p_z,B_\star)=\sqrt{p_z^2c^2+m_e^2c^4(1+2\nu B_\star)}$
with $\nu=n_\text{L}+\tfrac{1}{2}+\sigma$, where $\nu$ is the quantum number of the
Landau levels, $n_\text{L}$ is a non-negative integer number, $p_z$\,$=$\,$\hbar k_z$
is electron momentum along the magnetic field, $\sigma$\,$=$\,$\pm1/2$ represents the
spin degrees of freedom, $m_\text{e}$ is the electron mass, $c$ is the speed of light,
and $B_\star\equiv B/B_\text{c}$ (see, \textit{e.g.}, Ref.~\cite{Aggarwal(2021)}).
It is customary to quantify the magnetic field strength
relative to the critical value, $B_\text{c}=m_e^2c^3/e\hbar\simeq4.41\times10^{13}$\,G,
at which the cyclotron energy of electrons reaches its rest mass energy. The electron
number density, $n_\text{e}$, energy density, $\mathcal{E}_\text{e}$, and pressure,
$P_\text{e}$, are given by
\begin{eqnarray}
n_\text{e} &=& \frac{2B_\star}{(2\pi)^2\lambda_\text{e}^3}\sum_{\nu=0}^{\nu_\text{max}}g_\nu x_\text{e}(\nu,B_\star),\\
\mathcal{E}_\text{e} &=& \frac{B_\star m_\text{e}c^2}{(2\pi)^2\lambda_\text{e}^3}\sum_{\nu=0}^{\nu_\text{max}}g_\nu(1+2\nu B_\star)\,\psi_+\biggl[\frac{x_\text{e}(\nu,B_\star)}{\sqrt{1+2\nu B_\star}}\biggr],\label{Eq:n_e}\\
P_\text{e} &=& \frac{B_\star m_\text{e}c^2}{(2\pi)^2\lambda_\text{e}^3}\sum_{\nu=0}^{\nu_\text{max}}g_\nu(1+2\nu B_\star)\,\psi_-\biggl[\frac{x_\text{e}(\nu,B_\star)}{\sqrt{1+2\nu B_\star}}\biggr],
\end{eqnarray}
where $\lambda_\text{e}$\,$=$\,$\hbar/m_\text{e}c$ is the electron Compton wavelength,
$g_\nu$ represents the degeneracy ($g_\nu$\,$=$\,1 for $\nu$\,$=$\,0 and 2 for $\nu$\,$\ge$\,1),
$x_\text{e}(\nu,B_\star)$\,$=$\,$\sqrt{\gamma_e^2-(1+2\nu B_\star)}$ with $\gamma_\text{e}$\,$=$\,$\mu_\text{e}/m_\text{e}c^2$,
the electron chemcal potential relative to the electron rest mass energy, $\psi_\pm[x]=
x\sqrt{1+x^2}\pm\ln\bigl(x+\sqrt{1+x^2}\bigr)$, and $\nu_\text{max}$ denotes the highest
occupied Landau level that satisfies $E_{\nu_\text{max}}(p_z$\,$=$\,$0,B_\star)\le\mu_\text{e}$.

To determine the outer crust composition under a magnetic field, we employ the method
of Baym, Pethick, and Sutherland \cite{BPS(1971)} (which we call the BPS method below).
In the BPS method, the Gibbs energy per nucleon, $g=(\mathcal{E}_\text{WS}+P)/n$, is
minimized for a given pressure $P$ and a magnetic field strength $B$, where $\mathcal{E}_\text{WS}$
is the energy density of the Wigner-Seitz cell and $n$ is the average baryon (nucleon)
number density. The Gibbs energy per nucleon reads\footnote{We neglect a small contribution
from quantum zero-point motion of ions, which is typically a few orders of magnitude smaller
than $\mathcal{E}_\text{L}$ and its magnetic field dependence is known to be saturated
\cite{Baiko(2009)}. Thus, we can safely neglect this term.}
\begin{equation}
g(A,Z;P,B) = \frac{M'(A,Z)c^2}{A} + \frac{Z}{A}\biggl(\frac{4\mathcal{E}_\text{L}}{3n_\text{e}}+\mu_\text{e}\biggr),
\label{Eq:Gibbs}
\end{equation}
where $M'(A,Z)$ is the mass of an atomic nucleus with a mass number $A$ and a proton
number $Z$. We adopt the latest experimental data, AME2020 \cite{AME2020-I,AME2020-II},
whenever available. Note that in an ordinary outer crust all nuclei are pressure ionized
and the nuclear masses (not the atomic ones) should be used, correcting the tabulated
experimental atomic masses, $M(A,Z)$, as $M'(A,Z)c^2=M(A,Z)c^2-Zm_\text{e}c^2+B_\text{e}(Z)$,
where $B_\text{e}(Z)$ denotes the binding energy of electrons\footnote{$B_\text{e}$
is parametrized as a function of $Z$ as $B_\text{e}(Z)=1.44381\times10^{-5}Z^{2.39}
+1.55468\times10^{-12}Z^{5.35}$, in units of MeV \cite{Lunney(2003)}.}. When experimental
masses are not available, we adopt those predicted by a theoretical mass model. We employ
two variants of a microscopic mass model based on the Hartree-Fock-Bogoliubov (HFB) theory
(HFB-27 \cite{HFB27} and HFB-32 \cite{HFB30-32}), the most accurate mass model to date
which employs the Bayesian Machine Learning (BML) \cite{BML}, and a phenomenological
mass model (KTUY \cite{KTUY}). We note that the pressure ionization may be hindered
in the presence of a strong magnetic field, which may introduce certain corrections
to the atomic masses. In this work we disregard the effects of a magnetic field on
nuclear masses, as in Ref.~\cite{Chamel(2012)}. As will be discussed below, inclusion
of the latter effects on nuclear masses will strengthen our finding. $\mathcal{E}_\text{L}$
is the lattice part of energy density, $\mathcal{E}_\text{L}=-C_\text{M}(4\pi/3)^{1/3}Z^2e^2n_\text{nucl}^{4/3}$,
where $C_\text{M}$ is the so-called Madelung constant specific to the lattice. According
to the values of $C_\text{M}$ \cite{Baiko(2001)}\footnote{$C_\text{M}=0.895929255682$
for a body-centered-cubic (BCC) lattice.}, a BCC lattice is favored. $n_\text{nucl}$
is the number density of nuclei, which is given by $n_\text{nucl}=n/A=n_\text{p}/Z=1/V_\text{WS}$,
where $n_\text{p}$ is the proton number density and $V_\text{WS}$ denotes the volume
of the Wigner-Seitz cell. The lattice energy density does not depend on the magnetic
field, in accordance with the Bohr-van Leeuwen theorem \cite{VanVleck(1932)}. For given
$P$ and $B$, the Gibbs energy \eqref{Eq:Gibbs} is computed for all possible $A$ and $Z$,
and we regard the minimum one as the equiliburium composition. Thus, each layer of the
outer crust is assumed to be composed of a single kind of nuclei forming a perfect BCC
lattice, as in Refs.~\cite{Basilico(2015),Chamel(2012)}.

\begin{figure} [t]
\includegraphics[width=\columnwidth]{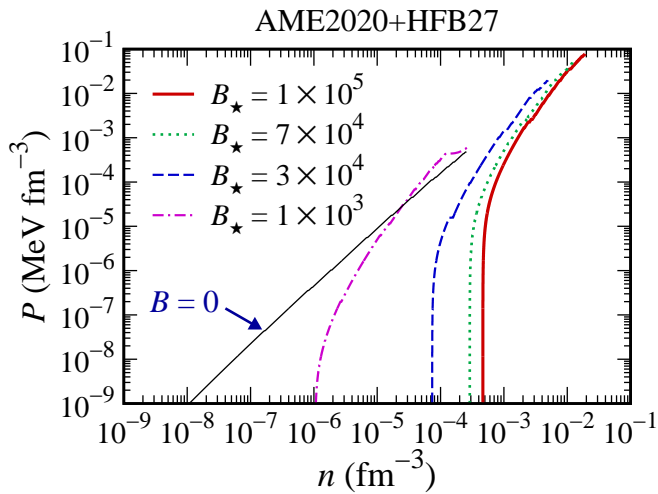}\vspace{-3mm}
\caption{
Equations of state. Pressure $P$ is plotted as a function of
average nucleon number density $n$ in the outer crust of
strongly-magnetized neutron stars for representative magnetic
field strengths.
}\vspace{-3mm}
\label{Fig:EoS}
\end{figure}

\textit{Effects of a magnetic field on the equation of state (EoS).---}
Let us begin with discussion on the EoS, \textit{i.e.} pressure $P$ as a function
of density $n$, for various magnetic field strengths; see Fig.~\ref{Fig:EoS}.
EoS for $B=0$ is also exhibited by a solid line for comparison. From the figure,
it is visible that the $P$-$n$ curve drops more and more rapidly at a small $n$
region, as the magnetic field strength increases. It is caused by the enhanced
electron fraction due to the Landau-Rabi quantization of electron motion. Because
of the charge neutrality condition, $n_\text{p}$\,$=$\,$n_\text{e}$, the crust
composition is kept fixed, which makes the crustal matter almost incompressible,
for a wide range of pressure. Note that the EoS lines end at high pressure and
density, when a transition from the outer crust to the inner one occurs, \textit{i.e.}
neutrons start dripping out of nuclei. The neutron drip occurs when the neutron chemical
potential exceeds its rest mass energy. Thus, we consider that the neutron drip
occurs when the condition, $\min_{A,Z}\bigl[g(A,Z;P,B_\star)\bigr]=\mu_\text{n}\geq
m_\text{n}c^2$, is fulfilled, where the first equality follows from the $\beta$-equilibrium
condition \cite{SM}. An important observation is that the neutron-drip pressure and
density are significantly increased for $B_\star$\,$\gtrsim$\,10$^4$. For $B_\star$\,$\gtrsim$\,10$^4$,
density goes well beyond $n_\text{drip}(B=0)$\,$\approx$\,$2.6\times10^{-4}$\,fm$^{-3}$,
which is never accessible for smaller magnetic field strengths. We have discovered
really exotic nuclear species in this high density region.

\begin{figure*} [t]
\includegraphics[width=\textwidth]{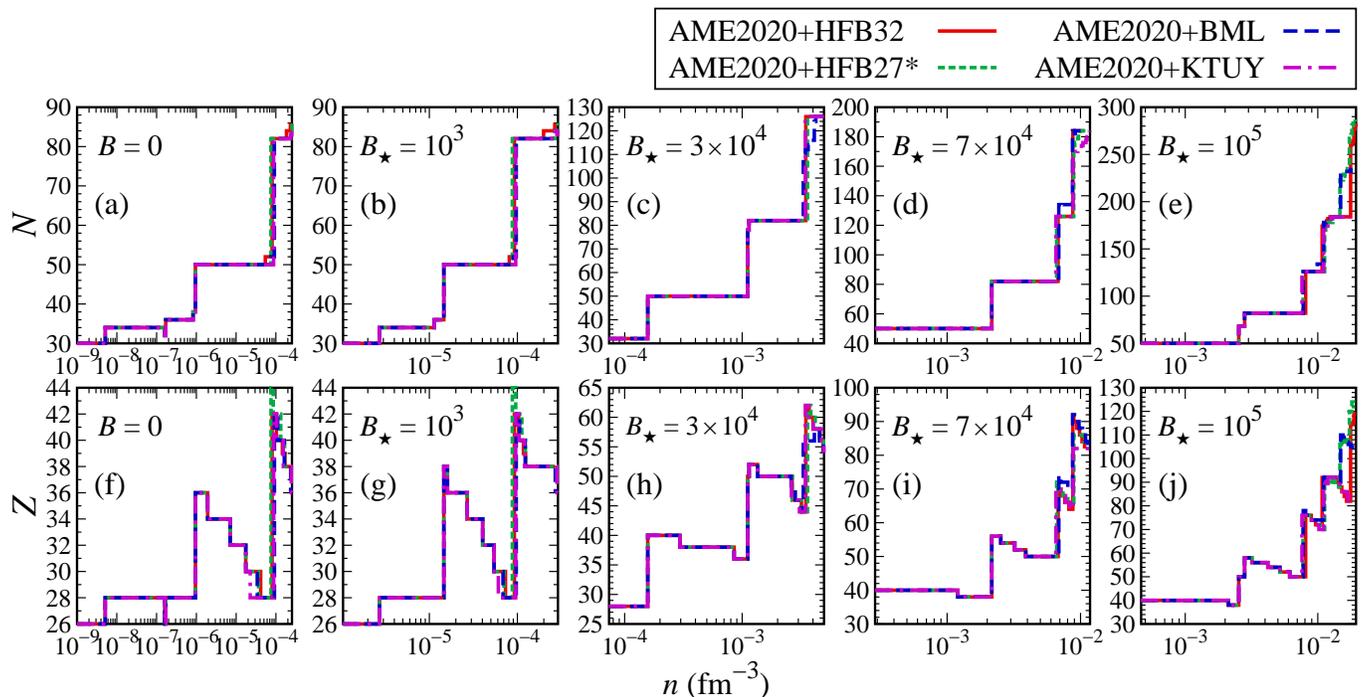}\vspace{-4mm}
\caption{
The outer-crust compositions calculated with the latest experimental masses,
AME2020 \cite{AME2020-I,AME2020-II}, supplemented with various theoretical mass models
(HFB-27 \cite{HFB27}, HFB-32 \cite{HFB30-32}, BML \cite{BML}, and KTUY \cite{KTUY})
are shown for representative magnetic field strengths. In upper (lower) panels, the
neutron (proton) number is shown as a function of average baryon (nucleon) number
density $n$. From left to right panels, the magnetic field strength corresponds to:
$B_\star=0$ (\textit{i.e.} without magnetic field), $10^3$ ($4.4\times10^{16}$\,G),
$3\times10^4$ ($1.3\times10^{18}$\,G), $7\times10^4$ ($3.1\times10^{18}$\,G), and
$10^5$ ($4.4\times10^{18}$\,G).
}\vspace{-3mm}
\label{Fig:NZ}
\end{figure*}

\textit{The finding: Emergence of $N$-rich heavy and superheavy nuclei.---}
The resulting outer crust compositions obtained with various mass models are presented
in Fig.~\ref{Fig:NZ} for representative magnetic field strengths. In the upper (lower)
panels, the neutron (proton) number is shown as a function of density. The composition
calculated without the magnetic field ($B=0$) is also shown in (a) and (f) for comparison.
For $B_\star$\,$\lesssim$\,10$^3$, nuclides that appear in the outer crust are not so
much dependent on the magnetic field strength and the emerging atomic number is, at most,
$Z=42$, consistent with earlier studies \cite{Chamel(2012),Basilico(2015)}. In this region,
the stabilization effects of neutron magic numbers $N=50$ and 82 and that of protons
$Z$\,$=$\,28 are pronounced. In stark contrast, we find substantial dependence of the
compositions on the magnetic field strength for $B_\star$\,$\gtrsim$\,10$^4$. At
around $B_\star$\,$=$\,3\,$\times$\,10$^4$ [(c) and (h)], nuclei with $Z$\,$>$\,42
emerge, which were absent for smaller $B$, where the next neutron magic number,
$N$\,$=$\,126, is reached at a bottom layer of the outer crust. At around $B_\star$\,$=$\,7\,$\times$10$^4$
[(d) and (i)], even more heavier nuclei $82\lesssim Z\lesssim92$ with $N$\,$=$\,184
show up. The latter two neutron magic numbers, $N$\,$=$\,126 and 184 with
$54$\,$\lesssim$\,$Z$\,$\lesssim$\,92 have been overlooked in preceding studies
\cite{Chamel(2012),Basilico(2015),Parmar(2022)}.

It is not the end of the story. Interestingly enough, at around $B_\star$\,$=$\,10$^5$,
the microscopic HFB-27 and HFB-32 mass models predict the existence of SHN with $Z$\,$>$\,104
including the unknown elements, 119, 120, 122, and 124, with a significantly large neutron
number $N$\,$\approx$\,260--287. Note that a naive counting of single-particle
orbitals in a spherical Woods-Saxon potential with the spin-orbit coupling gives us
$N$\,$=$\,258 as the next magic number after $N$\,$=$\,184. It explains a huge jump
from $N$\,$=$\,184 to 260--287 in the equilibrium compositions, observed in Fig.~\ref{Fig1}
and Fig.~\ref{Fig:NZ}(e). Intriguingly, the BML mass model captures $N$\,$=$\,184 magicity
in the lead region, while the next magic number around $N$\,$=$\,258 in the superheavy
region is not encoded, resulting in smaller mass and atomic numbers for $B_\star$\,$=$\,10$^5$
as compared to those of HFB mass models. Though it is not clear in Fig.~\ref{Fig:NZ},
but SHN do not appear for KTUY at $B_\star$\,$=$\,10$^5$. We have confirmed that SHN with
$Z$\,$=$\,104, 106, and 108 with $N$\,$=$\,184 and 200 appear also for KTUY at a slightly
larger $B$ value, $B_\star$\,$=$\,1.4\,$\times$\,10$^5$. Thus, our results are quite robust,
at least qualitatively, showing slight quantitative model dependence only in bottom layers
of the outer crust, where very $N$-rich nuclei come into play. We point out that the maximum
atomic number $Z_\text{max}$ involved in the available theoretical mass tables depends on
the models. For instance, $Z_\text{max}$ is 124, 120, 132, and 130 for HFB-27 \cite{HFB27},
HFB-32 \cite{HFB30-32}, BML \cite{BML}, and KTUY \cite{KTUY}, respectively, while some other
mass models do not include, \textit{e.g.}, $Z$\,$>$\,110 which limits applicability. We stress
here that to obtain correct equilibrium compositions under a superstrong magnetic field, it is
essential to include all possible $A$ and $Z$, without any biased exclusion of mass and/or
neutron numbers. (See Supplemental Material \cite{SM} for complete lists of the compositions.)

\textit{Discussion.---}One may claim that the predicted magnetic field strength
$B_\star$\,$\gtrsim$\,10$^5$ (\textit{i.e.} $B$\,$\gtrsim$\,4\,$\times$\,10$^{18}$\,G)
for the emergence of extremely $N$-rich SHN is too strong that is out of ream of existing
magnetars. However, there exist theoretical arguments that allow existence of magnetars
with $B\approx10^{18}$\,G \cite{Potekhin(1996),Potekhin(1999),Broderick(2000),Ventura(2001)},
and the possibility should not be ruled out. Moreover, we expect that the effects of a magnetic
field on nuclear structure, which has been neglected in this work, will strengthen the finding.
Namely, in Ref.~\cite{Basilico(2015)}, fully self-consistent CDFT calculations were performed
for Fe to Zn, showing that nuclear binding energy actually increases with the magnetic field
strength and it becomes larger for higher $Z$. We anticipate that if a strong magnetic field
does stabilize SHN as well, they could emerge at lower magnetic field strengths than
$B_\star$\,$\approx$\,10$^5$, thus, it should be regarded as an upper bound for the
critical magnetic field for the emergence of SHN.

We note, however, that the situation is not so obvious. First, nuclear mass models for
the ground states involve effects of pairing correlations, which must vanish above a
certain critical magnetic field ($B$\,$\gtrsim$\,10$^{17}$\,G) which breaks spin-singlet
Cooper pairs \cite{Stein(2016)_matter}. Second, it is known that a magnetic field affects
nuclear structure above $B$\,$\gtrsim$\,10$^{17}$\,G at which spin-dependent shifts of
single-particle energy levels become on the order of MeV \cite{Arteaga(2011),Stein(2016)_C-O-Ne}.
When such sifts of energy levels change their ordering or shell structure, nuclear deformation
as well as masses could be changed, which would affect the equilibrium composition. In addition,
because of the different signs of the intrinsic magnetic moment of neutrons and protons
($g_\text{n}$\,$=$\,$-$3.8263 and $g_\text{p}$\,$=$\,5.5856), time-odd currents rotate in
opposite directions about the magnetic field axis \cite{Stein(2016)_C-O-Ne}. The aforementioned
effects are expected to be larger for heavier nuclei, because of the higher level density around
the Fermi level and contributions of orbitals with larger angular momenta. To draw a quantitative
conclusion, it is thus highly desired to extend the works of Refs.~\cite{Arteaga(2011),Basilico(2015),
Stein(2016)_C-O-Ne} based on microscopic self-consistent calculations to a wider mass region,
including extremely $N$-rich SHN, under superstrong magnetic fields.

\begin{figure} [t]
\includegraphics[width=\columnwidth]{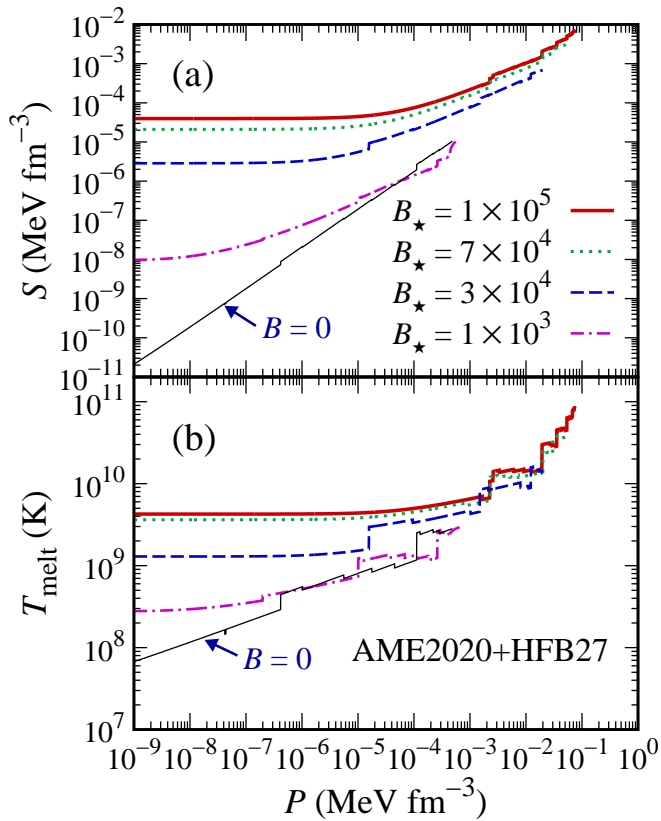}\vspace{-3mm}
\caption{
(a) Effective shear modulus $S$ and (b) melting temperature $T_\text{melt}$
for the outer crust of strongly-magnetized neutron stars are shown as
a function of pressure $P$ for representative magnetic field strengths
\cite{S_of_Chamel(2012)}.
}
\label{Fig:S_and_Tmelt}
\end{figure}

One may expect that the emergence of SHN may have a significant impact on the crustal
properties. As prominent examples we present in Figs.~\ref{Fig:S_and_Tmelt}(a) and
\ref{Fig:S_and_Tmelt}(b), respectively, the effective shear modulus of the outer crust,
$S$\,$=$\,0.1194\,$n_\text{nucl}Z^2e^2/R$ \cite{Ogata(1990),Chamel(2012)}, where
$R$\,$=$\,$(3/4\pi n_N)^{1/3}$ is radius of the Wigner-Seitz cell, and the melting
(or crystallizing) temperature, $T_\text{melt}$\,$=$\,$Z^2e^2/(k_\text{B}\Gamma R)$
\cite{Fantina(2020)}, where $k_\text{B}$ is the Boltzmann constant and $\Gamma$\,$=$\,175
is the Coulomb coupling parameter at melting, as a function of pressure for representative
magnetic field strengths. From the figure we find that both shear modulus and effective
temperature increase dramatically with the magnetic field strength. It is associated with
the increase of density $n$ for a given pressure $P$, as shown in Fig.~\ref{Fig:EoS}.
The abrupt changes of $S$ and $T_\text{melt}$ as a function of $P$ correspond to the
change of the compositions, since they are in the strongly quantized regime (\textit{i.e.}
$\nu_\text{max}$\,$=$\,0) for almost all pressure \cite{SM}. Because the neutron drip
pressure substantially increases with the magnetic field strengths $B$, the maximum
values of the effective shear modulus and the melting temperature increase with $B$.
The orders of magnitude increase of the effective shear modulus implies that the outer
crust of strongly-magnetized neutron stars is far more rigid than previously thought.
It would bring new insight into physics of quasi-periodic oscillations (QPO), giant flares
and starquakes, as well as a continuous gravitational wave emission from mountains at the
surface \cite{Horowitz(2009)}, and so forth. Further, the substantial increase of the melting
temperature may affect thermal evolution of strongly-magnetized proto-neutron stars. Since
the changes of crustal properties are so large, they could have detectable consequences
if such a strongly-magnetized neutron star is born.

\textit{Conclusions.---}In this work, we have systematically investigated outer crust
compositions for a wide range of pressures (densities) and magnetic field strengths,
adopting the latest experimental masses supplemented with various mass models. By exploring
the optimal equilibrium composition of nuclei from the entire nuclear chart, we have found,
for $B$\,$\gtrsim$\,10$^{18}$\,G, the emergence of neutron-rich heavy nuclei, which are much
heavier than those found in preceding studies, showing a clear manifestation of neutron magic
numbers, $N$\,$=$\,50, 82, 126, as well as 184. Moreover, for $B$\,$\gtrsim$\,4\,$\times$\,10$^{18}$\,G,
we have found that superheavy nuclei (SHN) with $Z$\,$\ge$\,104 naturally emerge as equilibrium
compositions that minimize the Gibbs energy. We have demonstrated that crustal properties, such
as shear modulus and melting temperature, are dramatically increased by orders of magnitude
by the magnetic field. As a consequence, \textit{e.g.}, the breaking strain of the crust may
become substantially larger than that without magnetic field effects. It results in a larger
maximum ellipticity which may be detected as continuous gravitational waves from mountains
at the neutron star surface, if such a superstrong magnetar exists. Such observations would imply
the existence of SHN, giving important feedback to microscopic theories.

The finding will open a new exciting research area, building a bridge between studies of
SHN and neutron stars, leaving a lot of open questions, \textit{e.g.}:
\textit{i) Crustal properties of strongly-magnetized neutron stars.} Our calculations indicate
that not only the crust compositions, but also the crustal properties are substantially altered
by the magnetic field. Its consequence, such as magnitude and frequency of crust oscillation
modes, the maximum height of mountains and gravitational signals, thermal and electrical
conductivities, as well as neutron star structure, should be reexamined.
\textit{ii) Effects of the magnetic field on nuclear structure.} In this work, we have found
the emergence of extremely neutron-rich SHN with $N$\,$\approx$\,260--287 for $B$\,$\gtrsim$\,4\,$\times$\,10$^{18}$\,G,
while the magnetic field effects on nuclear structure were neglected. To draw a quantitative
conclusion on the neutron and proton numbers of SHN and the critical magnetic field strength
for their appearance, fully microscopic, self-consistent calculations are necessary for the
entire nuclear chart, until the neutron drip line of SHN, for a wide range of magnetic field
strengths. We mention here that in Ref.~\cite{Agbemava(2021)} axially-symmetric CDFT calculations
were carried out including ``hyperheavy'' nuclei with $Z$\,$=$\,126--210, where existence of toroidal
nuclei has been advocated. Such calculations should also be achieved under superstrong magnetic
fields to judge if they could exist in strongly-magnetized neutron stars.
\textit{iii) Properties of the inner crust.} For the strongest magnetic field neutrons start
dripping out of SHN. Thus, the inner crust may also contain SHN under a superstrong magnetic field,
before they transform into inhomogeneous crystalline structure (the so-called pasta phase \cite{Caplan(2017)}).
In such a case, properties of dripped neutrons are most accurately described with the band theory
of solids (see, Refs.~\cite{Chamel(2017),Kashiwaba(2019),Sekizawa(2022)}, and references therein)
which has never been applied taking into account the magnetic field effects. How structure and
properties of the inner crust are affected by our finding is also an open question.

\begin{acknowledgments}
We are grateful Stephane Goriely for providing all the HFB mass tables currently available (from HFB-01 to HFB-32),
which were useful to conduct this study.
We are also thankful to Kei Iida and Takashi Nakatsukasa for providing useful comments on this Letter.
\end{acknowledgments}

\appendix
\clearpage
\begin{widetext}
\section{\large Supplemental Online Material for:\\
Possible Existence of Extremely Neutron-Rich Superheavy Nuclei in Neutron Star Crusts\\
Under a Superstrong Magnetic Field}

\begin{center}
Kazuyuki Sekizawa$^{1,2,*}$ and Kentaro Kaba$^1$\\
{\small
$^1$\textit{Department of Physics, School of Science, Tokyo Institute of Technology, Tokyo 152-8551, Japan}\\
$^2$\textit{Nuclear Physics Division, Center for Computational Sciences, University of Tsukuba, Ibaraki 305-8577, Japan}\\
$^3$\textit{RIKEN Nishina Center, Saitama 351-0198, Japan}\\
}

\vspace{5mm}
\begin{minipage}{0.78\textwidth}
{\small
In this Supplemental Material, we provide supplementary explanations that may help to understand
the effects of strong magnetic fields on the outer crust compositions, the main results of the paper.
We also present numerical tables showing various quantities obtained with four mass models used
(HFB-27, HFB-32, BML, and KTUY) for a full range of pressures (densities) and magnetic field strengths.
}
\end{minipage}
\end{center}
\end{widetext}

\section{On the neutron drip}

By definition, there follows a thermodynamic relation,
\begin{equation}
G \equiv gn = \mu_\text{n}n_\text{n} + \mu_\text{p}n_\text{p} + \mu_\text{e}n_\text{e},
\label{SMEq:G}
\end{equation}
where $G$ denotes the Gibbs energy ($g$ is the Gibbs energy per nucleon),
$n$ is the average baryon (nucleon) number density, $\mu_i$ and $n_i$ are the chemical
potential and the average number density of particle $i$ ($=$\,n, p, or e), respectively.
Because of the charge neutrality condition, $n_\text{p}$\,$=$\,$n_\text{e}$, and the
$\beta$-equilibrium condition, $\mu_\text{n}=\mu_\text{p}+\mu_\text{e}$, the
right hand side of Eq.~\eqref{SMEq:G} can be written as
\begin{equation}
G = \mu_\text{n}n.
\end{equation}
It means that the chemical potential of neutrons is equal to the Gibbs energy per nucleon,
\textit{i.e.} $\mu_\text{n}$\,$=$\,$g$, that is, the neutron drip pressure $P_\text{drip}$
is defined by the condition, $g(P_\text{drip})$\,$=$\,$\mu_\text{n}$\,$=$\,$m_\text{n}c^2$.
We thus performed calculations until the condition,
\begin{equation}
\min_{A,Z}\bigl[ g(A,Z; P,B) \bigr] \ge m_\text{n}c^2,
\end{equation}
is fulfilled for all sets of the mass and atomic numbers $(A,Z)$.

\begin{figure} [t]
\includegraphics[width=0.95\columnwidth]{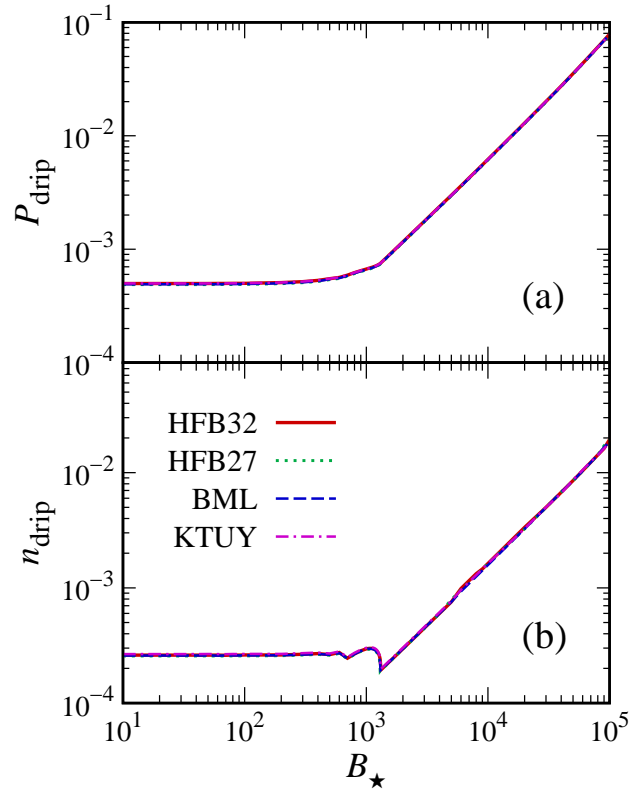}\vspace{-3mm}
\caption{
(a) Neutron-drip pressure $P_\text{drip}$ and (b) density $n_\text{drip}$
for strongly-magnetized neutron stars are shown as a function of the magnetic
field strength $B_\star$\,$=$\,$B/B_\text{c}$ with $B_\text{c}\simeq4.41\times10^{13}$\,G.
}\vspace{-3mm}
\label{Fig:P_drip}
\end{figure}

In Figs.~\ref{Fig:P_drip}(a) and \ref{Fig:P_drip}(b), we show the resulting neutron
drip pressure and density, $P_\text{drip}$ and $n_\text{drip}$, respectively, as a
function of the magnetic field strength $B_\star$\,$=$\,$B/B_\text{c}$ with $B_\text{c}
\simeq4.41\times10^{13}$\,G. From the figure, we see that $P_\text{drip}$ and $n_\text{drip}$
are almost independent from the mass models used. Also, it is visible that $P_\text{drip}$
and $n_\text{drip}$ are hardly affected for $B_\star$\,$\lesssim$\,10$^3$. As the magnetic
field strength increases, $P_\text{drip}$ and $n_\text{drip}$ start increasing nearly linearly
for $B_\star$\,$\gtrsim$\,10$^3$. A kink in between $B_\star$\,$=$\,1000 and 2000 is associated
with a transition to the strongly-quantizing regime where $\nu_\text{max}$\,$=$\,1\,$\rightarrow$\,0
takes place. The observed increase of the neutron drip density is the key to pave a way
to the region where very neutron-rich heavy and superheavy nuclei emerge, as we discussed
in the main texts.

\section{On the electron fraction}

It is important to understand why the electron fraction is supressed due to the
presence of a strong magnetic field. To see this, let us consider the strongly-quantizing
regime where only the lowest Landau level is occupied, \textit{i.e.} $\nu_\text{max}$\,$=$\,0.
In this case, the expression for the electron number density \eqref{Eq:n_e} is reduced to
\begin{equation}
n_\text{e} = \frac{2B_\star}{(2\pi)^2\lambda_\text{e}^3}\sqrt{\frac{\mu_\text{e}}{m_\text{e}c^2}-1}.
\label{SMEq:n_e}
\end{equation}
It is apparent that the electron number density $n_\text{e}$ is proportional to the
magnetic field strength, leading to enhancement of the electron fraction. Thus, for
a given pressure $P$, which has corresponding density $n$, the proton number density
is enhanced via the charge neutrality condition, $n_\text{p}$\,$=$\,$n_\text{e}$,
resulting in the reduction of neutron richness.

\begin{figure} [t]
\includegraphics[width=\columnwidth]{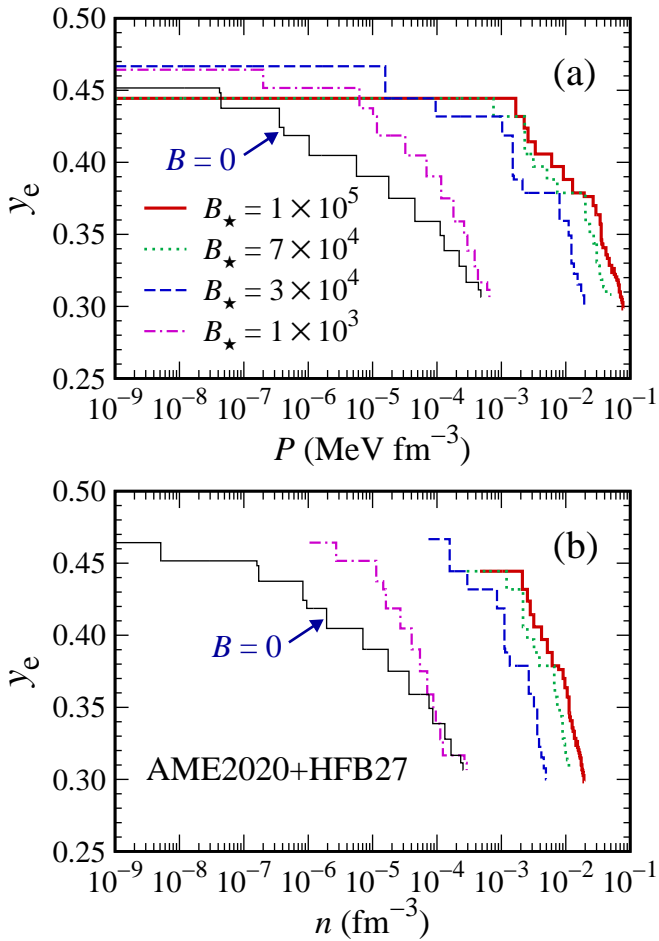}\vspace{-3mm}
\caption{
Electron fraction $y_\text{e}$\,$=$\,$Z/A$ in the outer crust of strongly-magnetized
neutron stars is shown as a function of pressure (a) and density (b) for
representative magnetic field strengths.
}\vspace{-3mm}
\label{Fig:y_e}
\end{figure}

In Fig.~\ref{Fig:y_e}(a), we show the calculated electron fraction $y_\text{e}$\,$=$\,$Z/A$
as a function of pressure $P$ for representative magnetic field strengths. The case
without magnetic field (\textit{i.e.} $B$\,$=$\,0) is also shown for comparison. A larger
electron fraction indicates a larger proton fraction and, thus, a less neutron-rich
composition. By comparing the results for $B$\,$=$\,0 (thin solid line) with $B_\star$\,$=$\,10$^3$
(dashed line) and 3\,$\times$\,10$^4$ (dash-dotted line) for $P$\,$<$\,$P_\text{drip}(B$\,$=$\,0$)$,
one finds an enhancement of the electron fraction due to the presence of the magnetic
field. On the other hand, we find that $y_\text{e}$ is actually smaller for
$B_\star$\,$=$\,7$\times$\,10$^4$ (dotted line) and $10^5$ (solid line) as compared
to the $B$\,$=$\,0 case for $P$\,$\lesssim$\,4\,$\times$\,10$^{-8}$. It is because of
the fact that density $n$ is much higher under such strong magnetic fields, see
Fig.~\ref{Fig:y_e}(b). In the figure, the same results are plotted as in Fig.~\ref{Fig:y_e}(a),
but the horizontal axis is replaced with density $n$. Because of the higher density,
neutron-rich heavy nuclei with larger neutron magic numbers $N$\,$=$\,50 and 126 emerge,
explaining a bit smaller value of $y_\text{e}$ for $P$\,$\lesssim$\,4\,$\times$\,10$^{-8}$.
As is evident from Fig.~\ref{Fig:y_e}(b), for $B_\star$\,$\gtrsim$\,10$^4$ density can
reach much higher values than $n_\text{drip}(B$\,$=$\,0$)$, where really exotic nuclei,
including extremely neutron-rich superheavy nuclei, emerge.

\section{Crust compositions}
\vspace{-2mm}

\begin{figure} [t]
\includegraphics[width=\columnwidth]{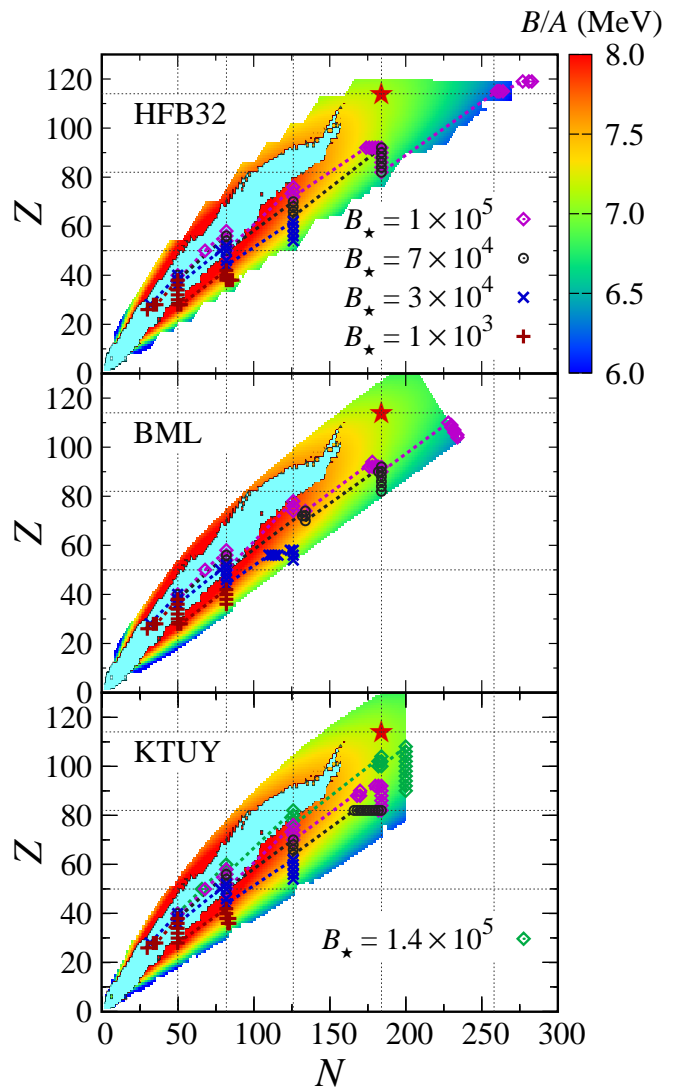}\vspace{-3mm}
\caption{
Same as Fig.~\ref{Fig1} (in the main texts), but calculated with other theoretical
mass models, HFB-32 \cite{HFB30-32} (top), BML \cite{BML}, and KTUY \cite{KTUY}.
For the results with the KTUY mass model, those for $B_\star=1.4\times10^5$ are also
shown, for which emergence of superheavy nuclei $Z=104$, 106 and 108 are observed.
}\vspace{-3mm}
\label{Fig:Compositions_SM}
\end{figure}

For completeness, we show in Fig.~\ref{Fig:Compositions_SM} the outer crust
compositions calculated with the latest experimental masses, AME2020 \cite{AME2020-I,AME2020-II},
supplemented with different mass models, HFB-32 \cite{HFB30-32} (top), BML \cite{BML} (middle),
and KTUY \cite{KTUY} (bottom). From Figs.~\ref{Fig1} and \ref{Fig:Compositions_SM}, we find
that the microscopic HFB mass models predicts a substantial stabilization effect around
$N=258$ in the superheavy region, whereas it is absent in the BML and KTUY mass models.
Apparently, BML and KTUY suffer from limited neutron numbers included in the mass table.
It is thus important to provide mass tables up to $Z$\,$\approx$\,130 or beyond, until
the drip lines.

\vspace{-3mm}
\section{Numerical tables}
\vspace{-2mm}

Below we provide numerical tables obtained in this study.



\begin{table*}[htbp]
\caption{
The results of BPS model calculations for $B=0$ with the experimental masses,
AME2020 \cite{AME2020-I,AME2020-II}, supplemented with the microscopic mass model,
HFB-27 \cite{HFB27}. In the first to fourth columns, the outer crust composition
[\textit{i.e.}, nuclides, $^A$X, where X stands for the corresponding symbol of the
element, proton, neutron, and mass numbers, $Z$, $N$, $A$, are presented, respectively.
In the fifth to sixth columns, the corresponding electron fraction, $y_\text{e}=Z/A$,
and $N/Z$ ratio are shown, respectively. In the seventh (eighth) column, the maximum
density $n_\text{max}$ (pressure $P_\text{max}$) are given in units of fm$^{-3}$
(MeV\,fm$^{-3}$), below which the corresponding nuclide exists. From the ninth
to eleventh columns, it shows: $R=(3/4\pi n_N)^{1/3}$, the radius of the Wigner-Seitz
cell in units of fm, $\hbar\omega_p$, energy of the plasma oscillation in units of MeV,
and $S=0.1194n_NZ^2e^2/R$, the effective shear modulus in units of MeV\,fm$^{-3}$, respectively.
}\vspace{-1mm}
\begin{center}
\begin{tabular*}{\textwidth}{@{\extracolsep{\fill}}ccccccccccc}
\hline
\hline
\multicolumn{11}{c}{AME2020\,+\,HFB27\; ($B=0$)}\\
\hline
Nuclides&$Z$&$N$&$A$&$y_e$&$N/Z$&$n_\text{max}$\,[fm$^{-3}$]&$P_\text{max}$\,[MeV/fm$^3$]&$R$\,[fm]&$\hbar\omega$\,[MeV]&$S$\,[MeV/fm$^3$]\\
\hline
\ce{^{ 56}Fe}&  26&  30&  56&0.46&1.15&$4.94\times10^{ -9}$&$3.37\times10^{-10}$&$1.39\times10^{  3}$&$8.98\times10^{ -4}$&$7.35\times10^{-12}$\\
\ce{^{ 62}Ni}&  28&  34&  62&0.45&1.21&$1.58\times10^{ -7}$&$4.16\times10^{ -8}$&$4.54\times10^{  2}$&$4.94\times10^{ -3}$&$7.55\times10^{-10}$\\
\ce{^{ 58}Fe}&  26&  32&  58&0.45&1.23&$1.65\times10^{ -7}$&$4.37\times10^{ -8}$&$4.38\times10^{  2}$&$5.01\times10^{ -3}$&$7.53\times10^{-10}$\\
\ce{^{ 64}Ni}&  28&  36&  64&0.44&1.29&$7.96\times10^{ -7}$&$3.53\times10^{ -7}$&$2.68\times10^{  2}$&$1.07\times10^{ -2}$&$6.25\times10^{ -9}$\\
\ce{^{ 66}Ni}&  28&  38&  66&0.42&1.36&$9.24\times10^{ -7}$&$4.14\times10^{ -7}$&$2.57\times10^{  2}$&$1.12\times10^{ -2}$&$7.33\times10^{ -9}$\\
\ce{^{ 86}Kr}&  36&  50&  86&0.42&1.39&$1.85\times10^{ -6}$&$1.02\times10^{ -6}$&$2.23\times10^{  2}$&$1.57\times10^{ -2}$&$2.15\times10^{ -8}$\\
\ce{^{ 84}Se}&  34&  50&  84&0.40&1.47&$6.77\times10^{ -6}$&$5.56\times10^{ -6}$&$1.43\times10^{  2}$&$2.90\times10^{ -2}$&$1.11\times10^{ -7}$\\
\ce{^{ 82}Ge}&  32&  50&  82&0.39&1.56&$1.67\times10^{ -5}$&$1.76\times10^{ -5}$&$1.05\times10^{  2}$&$4.38\times10^{ -2}$&$3.39\times10^{ -7}$\\
\ce{^{ 80}Zn}&  30&  50&  80&0.38&1.67&$3.49\times10^{ -5}$&$4.49\times10^{ -5}$&$8.18\times10^{  1}$&$6.09\times10^{ -2}$&$8.24\times10^{ -7}$\\
\ce{^{ 78}Ni}&  28&  50&  78&0.36&1.79&$7.17\times10^{ -5}$&$1.11\times10^{ -4}$&$6.38\times10^{  1}$&$8.36\times10^{ -2}$&$1.94\times10^{ -6}$\\
\ce{^{126}Ru}&  44&  82& 126&0.35&1.86&$8.28\times10^{ -5}$&$1.28\times10^{ -4}$&$7.13\times10^{  1}$&$8.74\times10^{ -2}$&$3.06\times10^{ -6}$\\
\ce{^{124}Mo}&  42&  82& 124&0.34&1.95&$1.27\times10^{ -4}$&$2.19\times10^{ -4}$&$6.15\times10^{  1}$&$1.05\times10^{ -1}$&$5.07\times10^{ -6}$\\
\ce{^{122}Zr}&  40&  82& 122&0.33&2.05&$1.58\times10^{ -4}$&$2.80\times10^{ -4}$&$5.68\times10^{  1}$&$1.14\times10^{ -1}$&$6.28\times10^{ -6}$\\
\ce{^{120}Sr}&  38&  82& 120&0.32&2.16&$2.27\times10^{ -4}$&$4.34\times10^{ -4}$&$5.01\times10^{  1}$&$1.31\times10^{ -1}$&$9.39\times10^{ -6}$\\
\ce{^{122}Sr}&  38&  84& 122&0.31&2.21&$2.47\times10^{ -4}$&$4.75\times10^{ -4}$&$4.90\times10^{  1}$&$1.35\times10^{ -1}$&$1.03\times10^{ -5}$\\
\ce{^{124}Sr}&  38&  86& 124&0.31&2.26&$2.55\times10^{ -4}$&$4.84\times10^{ -4}$&$4.88\times10^{  1}$&$1.35\times10^{ -1}$&$1.05\times10^{ -5}$\\
\hline
\hline
\end{tabular*}
\end{center}
\label{tab:HFB27-B_star=0}
\end{table*}

\begin{table*}[htbp]
\caption{
Smae as Table~\ref{tab:HFB27-B_star=0}, but for $B_\star=10^3$. Here and also in the tables below,
the highest occupied Landau level $\nu_\text{max}$ is also shown for $B$\,$\neq$\,0 cases.
}\vspace{-1mm}
\begin{center}
\begin{tabular*}{\textwidth}{@{\extracolsep{\fill}}cccccccccccc}
\hline
\hline
\multicolumn{12}{c}{AME2020\,+\,HFB27\; ($B_\star=10^3$)}\\
\hline
Nuclides&$Z$&$N$&$A$&$y_e$&$N/Z$&$n_\text{max}$\,[fm$^{-3}$]&$P_\text{max}$\,[MeV/fm$^3$]&$\nu_\text{max}$&$R$\,[fm]&$\hbar\omega$\,[MeV]&$S$\,[MeV/fm$^3$]\\
\hline
\ce{^{ 56}Fe}&  26&  30&  56&0.46&1.15&$2.62\times10^{ -6}$&$1.97\times10^{ -7}$&      0&$1.72\times10^{  2}$&$2.07\times10^{ -2}$&$3.16\times10^{ -8}$\\
\ce{^{ 62}Ni}&  28&  34&  62&0.45&1.21&$1.10\times10^{ -5}$&$6.18\times10^{ -6}$&      0&$1.10\times10^{  2}$&$4.12\times10^{ -2}$&$2.16\times10^{ -7}$\\
\ce{^{ 64}Ni}&  28&  36&  64&0.44&1.29&$1.42\times10^{ -5}$&$1.01\times10^{ -5}$&      0&$1.02\times10^{  2}$&$4.54\times10^{ -2}$&$2.93\times10^{ -7}$\\
\ce{^{ 88}Sr}&  38&  50&  88&0.43&1.32&$1.55\times10^{ -5}$&$1.16\times10^{ -5}$&      0&$1.11\times10^{  2}$&$4.68\times10^{ -2}$&$3.96\times10^{ -7}$\\
\ce{^{ 86}Kr}&  36&  50&  86&0.42&1.39&$2.60\times10^{ -5}$&$3.19\times10^{ -5}$&      0&$9.25\times10^{  1}$&$5.87\times10^{ -2}$&$7.28\times10^{ -7}$\\
\ce{^{ 84}Se}&  34&  50&  84&0.40&1.47&$3.88\times10^{ -5}$&$6.79\times10^{ -5}$&      0&$8.03\times10^{  1}$&$6.93\times10^{ -2}$&$1.14\times10^{ -6}$\\
\ce{^{ 82}Ge}&  32&  50&  82&0.39&1.56&$5.20\times10^{ -5}$&$1.15\times10^{ -4}$&      0&$7.22\times10^{  1}$&$7.75\times10^{ -2}$&$1.55\times10^{ -6}$\\
\ce{^{ 80}Zn}&  30&  50&  80&0.38&1.67&$6.71\times10^{ -5}$&$1.78\times10^{ -4}$&      0&$6.58\times10^{  1}$&$8.45\times10^{ -2}$&$1.97\times10^{ -6}$\\
\ce{^{ 78}Ni}&  28&  50&  78&0.36&1.79&$8.45\times10^{ -5}$&$2.60\times10^{ -4}$&      0&$6.04\times10^{  1}$&$9.08\times10^{ -2}$&$2.42\times10^{ -6}$\\
\ce{^{126}Ru}&  44&  82& 126&0.35&1.86&$9.25\times10^{ -5}$&$2.93\times10^{ -4}$&      0&$6.88\times10^{  1}$&$9.24\times10^{ -2}$&$3.55\times10^{ -6}$\\
\ce{^{124}Mo}&  42&  82& 124&0.34&1.95&$1.08\times10^{ -4}$&$3.80\times10^{ -4}$&      0&$6.49\times10^{  1}$&$9.70\times10^{ -2}$&$4.08\times10^{ -6}$\\
\ce{^{122}Zr}&  40&  82& 122&0.33&2.05&$1.19\times10^{ -4}$&$4.28\times10^{ -4}$&      0&$6.26\times10^{  1}$&$9.82\times10^{ -2}$&$4.27\times10^{ -6}$\\
\ce{^{120}Sr}&  38&  82& 120&0.32&2.16&$2.60\times10^{ -4}$&$5.94\times10^{ -4}$&  0,  1&$4.79\times10^{  1}$&$1.40\times10^{ -1}$&$1.12\times10^{ -5}$\\
\ce{^{122}Sr}&  38&  84& 122&0.31&2.21&$2.82\times10^{ -4}$&$6.44\times10^{ -4}$&      1&$4.69\times10^{  1}$&$1.44\times10^{ -1}$&$1.23\times10^{ -5}$\\
\ce{^{124}Sr}&  38&  86& 124&0.31&2.26&$2.89\times10^{ -4}$&$6.50\times10^{ -4}$&      1&$4.68\times10^{  1}$&$1.43\times10^{ -1}$&$1.24\times10^{ -5}$\\
\hline
\hline
\end{tabular*}
\end{center}
\label{tab:HFB27-B_star=1000}
\end{table*}

\begin{table*}[htbp]
\caption{
Smae as Tables~\ref{tab:HFB27-B_star=0} and \ref{tab:HFB27-B_star=1000}, but for $B_\star=3\times10^4$.
}\vspace{-1mm}
\begin{center}
\begin{tabular*}{\textwidth}{@{\extracolsep{\fill}}cccccccccccc}
\hline
\hline
\multicolumn{12}{c}{AME2020\,+\,HFB27\; ($B_\star=3\times10^4$)}\\
\hline
Nuclides&$Z$&$N$&$A$&$y_e$&$N/Z$&$n_\text{max}$\,[fm$^{-3}$]&$P_\text{max}$\,[MeV/fm$^3$]&$\nu_\text{max}$&$R$\,[fm]&$\hbar\omega$\,[MeV]&$S$\,[MeV/fm$^3$]\\
\hline
\ce{^{ 60}Ni}&  28&  32&  60&0.47&1.14&$1.37\times10^{ -4}$&$1.56\times10^{ -5}$&      0&$4.71\times10^{  1}$&$1.50\times10^{ -1}$&$6.54\times10^{ -6}$\\
\ce{^{ 90}Zr}&  40&  50&  90&0.44&1.25&$2.88\times10^{ -4}$&$9.43\times10^{ -5}$&      0&$4.21\times10^{  1}$&$2.08\times10^{ -1}$&$2.09\times10^{ -5}$\\
\ce{^{ 88}Sr}&  38&  50&  88&0.43&1.32&$8.27\times10^{ -4}$&$1.02\times10^{ -3}$&      0&$2.94\times10^{  1}$&$3.42\times10^{ -1}$&$7.94\times10^{ -5}$\\
\ce{^{ 86}Kr}&  36&  50&  86&0.42&1.39&$1.01\times10^{ -3}$&$1.48\times10^{ -3}$&      0&$2.73\times10^{  1}$&$3.67\times10^{ -1}$&$9.65\times10^{ -5}$\\
\ce{^{128}Sn}&  50&  78& 128&0.39&1.56&$1.13\times10^{ -3}$&$1.54\times10^{ -3}$&      0&$3.00\times10^{  1}$&$3.61\times10^{ -1}$&$1.26\times10^{ -4}$\\
\ce{^{134}Te}&  52&  82& 134&0.39&1.58&$1.31\times10^{ -3}$&$2.10\times10^{ -3}$&      0&$2.90\times10^{  1}$&$3.87\times10^{ -1}$&$1.57\times10^{ -4}$\\
\ce{^{132}Sn}&  50&  82& 132&0.38&1.64&$2.52\times10^{ -3}$&$7.90\times10^{ -3}$&      0&$2.32\times10^{  1}$&$5.23\times10^{ -1}$&$3.53\times10^{ -4}$\\
\ce{^{128}Pd}&  46&  82& 128&0.36&1.78&$3.10\times10^{ -3}$&$1.10\times10^{ -2}$&      0&$2.14\times10^{  1}$&$5.51\times10^{ -1}$&$4.11\times10^{ -4}$\\
\ce{^{126}Ru}&  44&  82& 126&0.35&1.86&$3.35\times10^{ -3}$&$1.21\times10^{ -2}$&      0&$2.08\times10^{  1}$&$5.56\times10^{ -1}$&$4.25\times10^{ -4}$\\
\ce{^{188}Sm}&  62& 126& 188&0.33&2.03&$3.74\times10^{ -3}$&$1.33\times10^{ -2}$&      0&$2.29\times10^{  1}$&$5.54\times10^{ -1}$&$5.74\times10^{ -4}$\\
\ce{^{186}Nd}&  60& 126& 186&0.32&2.10&$4.06\times10^{ -3}$&$1.51\times10^{ -2}$&      0&$2.22\times10^{  1}$&$5.65\times10^{ -1}$&$6.09\times10^{ -4}$\\
\ce{^{184}Ce}&  58& 126& 184&0.32&2.17&$4.42\times10^{ -3}$&$1.72\times10^{ -2}$&      0&$2.15\times10^{  1}$&$5.76\times10^{ -1}$&$6.47\times10^{ -4}$\\
\ce{^{182}Ba}&  56& 126& 182&0.31&2.25&$4.77\times10^{ -3}$&$1.92\times10^{ -2}$&      0&$2.09\times10^{  1}$&$5.84\times10^{ -1}$&$6.77\times10^{ -4}$\\
\ce{^{180}Xe}&  54& 126& 180&0.30&2.33&$4.96\times10^{ -3}$&$1.97\times10^{ -2}$&      0&$2.05\times10^{  1}$&$5.81\times10^{ -1}$&$6.73\times10^{ -4}$\\
\hline
\hline
\end{tabular*}
\end{center}
\label{tab:HFB27-B_star=30000}
\end{table*}

\begin{table*}[htbp]
\caption{
Smae as Tables~\ref{tab:HFB27-B_star=0}--\ref{tab:HFB27-B_star=30000}, but for $B_\star=7\times10^4$.
}\vspace{-1mm}
\begin{center}
\begin{tabular*}{\textwidth}{@{\extracolsep{\fill}}cccccccccccc}
\hline
\hline
\multicolumn{12}{c}{AME2020\,+\,HFB27\; ($B_\star=7\times10^4$)}\\
\hline
Nuclides&$Z$&$N$&$A$&$y_e$&$N/Z$&$n_\text{max}$\,[fm$^{-3}$]&$P_\text{max}$\,[MeV/fm$^3$]&$\nu_\text{max}$&$R$\,[fm]&$\hbar\omega$\,[MeV]&$S$\,[MeV/fm$^3$]\\
\hline
\ce{^{ 90}Zr}&  40&  50&  90&0.44&1.25&$1.17\times10^{ -3}$&$7.47\times10^{ -4}$&      0&$2.64\times10^{  1}$&$4.18\times10^{ -1}$&$1.36\times10^{ -4}$\\
\ce{^{ 88}Sr}&  38&  50&  88&0.43&1.32&$1.95\times10^{ -3}$&$2.28\times10^{ -3}$&      0&$2.21\times10^{  1}$&$5.25\times10^{ -1}$&$2.49\times10^{ -4}$\\
\ce{^{138}Ba}&  56&  82& 138&0.41&1.46&$2.47\times10^{ -3}$&$3.10\times10^{ -3}$&      0&$2.37\times10^{  1}$&$5.55\times10^{ -1}$&$4.07\times10^{ -4}$\\
\ce{^{136}Xe}&  54&  82& 136&0.40&1.52&$3.13\times10^{ -3}$&$5.05\times10^{ -3}$&      0&$2.18\times10^{  1}$&$6.12\times10^{ -1}$&$5.30\times10^{ -4}$\\
\ce{^{134}Te}&  52&  82& 134&0.39&1.58&$3.79\times10^{ -3}$&$7.30\times10^{ -3}$&      0&$2.04\times10^{  1}$&$6.57\times10^{ -1}$&$6.45\times10^{ -4}$\\
\ce{^{132}Sn}&  50&  82& 132&0.38&1.64&$6.21\times10^{ -3}$&$1.99\times10^{ -2}$&      0&$1.72\times10^{  1}$&$8.21\times10^{ -1}$&$1.18\times10^{ -3}$\\
\ce{^{198}Hf}&  72& 126& 198&0.36&1.75&$6.62\times10^{ -3}$&$2.01\times10^{ -2}$&      0&$1.93\times10^{  1}$&$8.14\times10^{ -1}$&$1.55\times10^{ -3}$\\
\ce{^{196}Yb}&  70& 126& 196&0.36&1.80&$7.22\times10^{ -3}$&$2.34\times10^{ -2}$&      0&$1.86\times10^{  1}$&$8.35\times10^{ -1}$&$1.67\times10^{ -3}$\\
\ce{^{194}Er}&  68& 126& 194&0.35&1.85&$7.85\times10^{ -3}$&$2.69\times10^{ -2}$&      0&$1.81\times10^{  1}$&$8.54\times10^{ -1}$&$1.78\times10^{ -3}$\\
\ce{^{192}Dy}&  66& 126& 192&0.34&1.91&$8.41\times10^{ -3}$&$3.00\times10^{ -2}$&      0&$1.76\times10^{  1}$&$8.67\times10^{ -1}$&$1.86\times10^{ -3}$\\
\ce{^{260}Rn}&  86& 174& 260&0.33&2.02&$8.95\times10^{ -3}$&$3.06\times10^{ -2}$&      0&$1.91\times10^{  1}$&$8.61\times10^{ -1}$&$2.30\times10^{ -3}$\\
\ce{^{262}Rn}&  86& 176& 262&0.33&2.05&$9.37\times10^{ -3}$&$3.31\times10^{ -2}$&      0&$1.88\times10^{  1}$&$8.73\times10^{ -1}$&$2.41\times10^{ -3}$\\
\ce{^{265}Rn}&  86& 179& 265&0.32&2.08&$9.52\times10^{ -3}$&$3.34\times10^{ -2}$&      0&$1.88\times10^{  1}$&$8.70\times10^{ -1}$&$2.43\times10^{ -3}$\\
\ce{^{268}Rn}&  86& 182& 268&0.32&2.12&$9.85\times10^{ -3}$&$3.51\times10^{ -2}$&      0&$1.86\times10^{  1}$&$8.76\times10^{ -1}$&$2.51\times10^{ -3}$\\
\ce{^{270}Rn}&  86& 184& 270&0.32&2.14&$1.03\times10^{ -2}$&$3.81\times10^{ -2}$&      0&$1.84\times10^{  1}$&$8.89\times10^{ -1}$&$2.63\times10^{ -3}$\\
\ce{^{268}Po}&  84& 184& 268&0.31&2.19&$1.10\times10^{ -2}$&$4.20\times10^{ -2}$&      0&$1.80\times10^{  1}$&$9.02\times10^{ -1}$&$2.76\times10^{ -3}$\\
\ce{^{266}Pb}&  82& 184& 266&0.31&2.24&$1.22\times10^{ -2}$&$5.13\times10^{ -2}$&      0&$1.73\times10^{  1}$&$9.37\times10^{ -1}$&$3.07\times10^{ -3}$\\
\hline
\hline
\end{tabular*}
\end{center}
\label{tab:HFB27-B_star=70000}
\end{table*}

\begin{table*}[htbp]
\caption{
Smae as Tables~\ref{tab:HFB27-B_star=0}--\ref{tab:HFB27-B_star=70000}, but for $B_\star=10^5$.
}\vspace{-1mm}
\begin{center}
\begin{tabular*}{\textwidth}{@{\extracolsep{\fill}}cccccccccccc}
\hline
\hline
\multicolumn{12}{c}{AME2020\,+\,HFB27\; ($B_\star=10^5$)}\\
\hline
Nuclides&$Z$&$N$&$A$&$y_e$&$N/Z$&$n_\text{max}$\,[fm$^{-3}$]&$P_\text{max}$\,[MeV/fm$^3$]&$\nu_\text{max}$&$R$\,[fm]&$\hbar\omega$\,[MeV]&$S$\,[MeV/fm$^3$]\\
\hline
\ce{^{ 90}Zr}&  40&  50&  90&0.44&1.25&$2.06\times10^{ -3}$&$1.66\times10^{ -3}$&      0&$2.18\times10^{  1}$&$5.55\times10^{ -1}$&$2.88\times10^{ -4}$\\
\ce{^{ 88}Sr}&  38&  50&  88&0.43&1.32&$2.40\times10^{ -3}$&$2.23\times10^{ -3}$&      0&$2.06\times10^{  1}$&$5.82\times10^{ -1}$&$3.28\times10^{ -4}$\\
\ce{^{118}Sn}&  50&  68& 118&0.42&1.36&$2.68\times10^{ -3}$&$2.57\times10^{ -3}$&      0&$2.19\times10^{  1}$&$6.04\times10^{ -1}$&$4.45\times10^{ -4}$\\
\ce{^{140}Ce}&  58&  82& 140&0.41&1.41&$3.12\times10^{ -3}$&$3.32\times10^{ -3}$&      0&$2.20\times10^{  1}$&$6.37\times10^{ -1}$&$5.85\times10^{ -4}$\\
\ce{^{138}Ba}&  56&  82& 138&0.41&1.46&$4.10\times10^{ -3}$&$5.98\times10^{ -3}$&      0&$2.00\times10^{  1}$&$7.15\times10^{ -1}$&$8.01\times10^{ -4}$\\
\ce{^{136}Xe}&  54&  82& 136&0.40&1.52&$5.05\times10^{ -3}$&$9.08\times10^{ -3}$&      0&$1.86\times10^{  1}$&$7.76\times10^{ -1}$&$1.00\times10^{ -3}$\\
\ce{^{134}Te}&  52&  82& 134&0.39&1.58&$5.98\times10^{ -3}$&$1.26\times10^{ -2}$&      0&$1.75\times10^{  1}$&$8.26\times10^{ -1}$&$1.19\times10^{ -3}$\\
\ce{^{132}Sn}&  50&  82& 132&0.38&1.64&$7.43\times10^{ -3}$&$1.92\times10^{ -2}$&      0&$1.62\times10^{  1}$&$8.98\times10^{ -1}$&$1.49\times10^{ -3}$\\
\ce{^{202}Os}&  76& 126& 202&0.38&1.66&$8.91\times10^{ -3}$&$2.63\times10^{ -2}$&      0&$1.75\times10^{  1}$&$9.77\times10^{ -1}$&$2.50\times10^{ -3}$\\
\ce{^{200} W}&  74& 126& 200&0.37&1.70&$9.47\times10^{ -3}$&$2.91\times10^{ -2}$&      0&$1.71\times10^{  1}$&$9.91\times10^{ -1}$&$2.60\times10^{ -3}$\\
\ce{^{198}Hf}&  72& 126& 198&0.36&1.75&$1.03\times10^{ -2}$&$3.38\times10^{ -2}$&      0&$1.66\times10^{  1}$&$1.02\times10^{  0}$&$2.80\times10^{ -3}$\\
\ce{^{196}Yb}&  70& 126& 196&0.36&1.80&$1.07\times10^{ -2}$&$3.51\times10^{ -2}$&      0&$1.64\times10^{  1}$&$1.02\times10^{  0}$&$2.81\times10^{ -3}$\\
\ce{^{260}Th}&  90& 170& 260&0.35&1.89&$1.13\times10^{ -2}$&$3.58\times10^{ -2}$&      0&$1.76\times10^{  1}$&$1.01\times10^{  0}$&$3.44\times10^{ -3}$\\
\ce{^{268} U}&  92& 176& 268&0.34&1.91&$1.16\times10^{ -2}$&$3.73\times10^{ -2}$&      0&$1.76\times10^{  1}$&$1.02\times10^{  0}$&$3.58\times10^{ -3}$\\
\ce{^{270} U}&  92& 178& 270&0.34&1.93&$1.23\times10^{ -2}$&$4.16\times10^{ -2}$&      0&$1.73\times10^{  1}$&$1.04\times10^{  0}$&$3.83\times10^{ -3}$\\
\ce{^{268}Th}&  90& 178& 268&0.34&1.98&$1.28\times10^{ -2}$&$4.38\times10^{ -2}$&      0&$1.71\times10^{  1}$&$1.04\times10^{  0}$&$3.89\times10^{ -3}$\\
\ce{^{276} U}&  92& 184& 276&0.33&2.00&$1.33\times10^{ -2}$&$4.69\times10^{ -2}$&      0&$1.70\times10^{  1}$&$1.06\times10^{  0}$&$4.12\times10^{ -3}$\\
\ce{^{274}Th}&  90& 184& 274&0.33&2.04&$1.41\times10^{ -2}$&$5.13\times10^{ -2}$&      0&$1.67\times10^{  1}$&$1.07\times10^{  0}$&$4.29\times10^{ -3}$\\
\ce{^{272}Ra}&  88& 184& 272&0.32&2.09&$1.45\times10^{ -2}$&$5.34\times10^{ -2}$&      0&$1.65\times10^{  1}$&$1.07\times10^{  0}$&$4.32\times10^{ -3}$\\
\ce{^{331}Hs}& \textbf{108}& 223& 331&0.33&2.06&$1.50\times10^{ -2}$&$5.61\times10^{ -2}$&      0&$1.74\times10^{  1}$&$1.10\times10^{  0}$&$5.21\times10^{ -3}$\\
\ce{^{329}Sg}& \textbf{106}& 223& 329&0.32&2.10&$1.54\times10^{ -2}$&$5.84\times10^{ -2}$&      0&$1.72\times10^{  1}$&$1.10\times10^{  0}$&$5.26\times10^{ -3}$\\
\ce{^{337}Hs}& \textbf{108}& 229& 337&0.32&2.12&$1.59\times10^{ -2}$&$6.14\times10^{ -2}$&      0&$1.72\times10^{  1}$&$1.11\times10^{  0}$&$5.50\times10^{ -3}$\\
\ce{^{333}Sg}& \textbf{106}& 227& 333&0.32&2.14&$1.63\times10^{ -2}$&$6.45\times10^{ -2}$&      0&$1.69\times10^{  1}$&$1.12\times10^{  0}$&$5.59\times10^{ -3}$\\
\ce{^{335}Sg}& \textbf{106}& 229& 335&0.32&2.16&$1.65\times10^{ -2}$&$6.51\times10^{ -2}$&      0&$1.69\times10^{  1}$&$1.12\times10^{  0}$&$5.63\times10^{ -3}$\\
\ce{^{343}Hs}& \textbf{108}& 235& 343&0.31&2.18&$1.69\times10^{ -2}$&$6.71\times10^{ -2}$&      0&$1.69\times10^{  1}$&$1.12\times10^{  0}$&$5.82\times10^{ -3}$\\
\ce{^{387}Ubn}& \textbf{120}& 267& 387&0.31&2.23&$1.75\times10^{ -2}$&$6.98\times10^{ -2}$&      0&$1.74\times10^{  1}$&$1.13\times10^{  0}$&$6.46\times10^{ -3}$\\
\ce{^{391}Ubn}& \textbf{120}& 271& 391&0.31&2.26&$1.79\times10^{ -2}$&$7.12\times10^{ -2}$&      0&$1.73\times10^{  1}$&$1.13\times10^{  0}$&$6.54\times10^{ -3}$\\
\ce{^{405}Ubq}& \textbf{124}& 281& 405&0.31&2.27&$1.82\times10^{ -2}$&$7.34\times10^{ -2}$&      0&$1.74\times10^{  1}$&$1.14\times10^{  0}$&$6.83\times10^{ -3}$\\
\ce{^{407}Ubq}& \textbf{124}& 283& 407&0.30&2.28&$1.84\times10^{ -2}$&$7.41\times10^{ -2}$&      0&$1.74\times10^{  1}$&$1.14\times10^{  0}$&$6.87\times10^{ -3}$\\
\ce{^{403}Ubb}& \textbf{122}& 281& 403&0.30&2.30&$1.88\times10^{ -2}$&$7.64\times10^{ -2}$&      0&$1.72\times10^{  1}$&$1.14\times10^{  0}$&$6.91\times10^{ -3}$\\
\ce{^{409}Ubb}& \textbf{122}& 287& 409&0.30&2.35&$1.92\times10^{ -2}$&$7.79\times10^{ -2}$&      0&$1.72\times10^{  1}$&$1.14\times10^{  0}$&$7.00\times10^{ -3}$\\
\hline
\hline
\end{tabular*}
\end{center}
\label{tab:HFB27-B_star=100000}
\end{table*}


\begin{table*}[htbp]
\caption{
Smae as Table~\ref{tab:HFB27-B_star=0} (\textit{i.e.} $B=0$), but calculated with the HFB-32 mass model \cite{HFB30-32}.
}\vspace{-1mm}
\begin{center}
\begin{tabular*}{\textwidth}{@{\extracolsep{\fill}}ccccccccccc}
\hline
\hline
\multicolumn{11}{c}{AME2020\,+\,HFB32\; ($B=0$)}\\
\hline
Nuclides&$Z$&$N$&$A$&$y_e$&$N/Z$&$n_\text{max}$\,[fm$^{-3}$]&$P_\text{max}$\,[MeV/fm$^3$]&$R$\,[fm]&$\hbar\omega_\text{p}$\,[MeV]&$S$\,[MeV/fm$^3$]\\
\hline
\ce{^{ 56}Fe}&  26&  30&  56&0.46&1.15&$4.94\times10^{ -9}$&$3.37\times10^{-10}$&$1.39\times10^{  3}$&$8.98\times10^{ -4}$&$7.35\times10^{-12}$\\
\ce{^{ 62}Ni}&  28&  34&  62&0.45&1.21&$1.58\times10^{ -7}$&$4.16\times10^{ -8}$&$4.54\times10^{  2}$&$4.94\times10^{ -3}$&$7.55\times10^{-10}$\\
\ce{^{ 58}Fe}&  26&  32&  58&0.45&1.23&$1.65\times10^{ -7}$&$4.37\times10^{ -8}$&$4.38\times10^{  2}$&$5.01\times10^{ -3}$&$7.53\times10^{-10}$\\
\ce{^{ 64}Ni}&  28&  36&  64&0.44&1.29&$7.96\times10^{ -7}$&$3.53\times10^{ -7}$&$2.68\times10^{  2}$&$1.07\times10^{ -2}$&$6.25\times10^{ -9}$\\
\ce{^{ 66}Ni}&  28&  38&  66&0.42&1.36&$9.24\times10^{ -7}$&$4.14\times10^{ -7}$&$2.57\times10^{  2}$&$1.12\times10^{ -2}$&$7.33\times10^{ -9}$\\
\ce{^{ 86}Kr}&  36&  50&  86&0.42&1.39&$1.85\times10^{ -6}$&$1.02\times10^{ -6}$&$2.23\times10^{  2}$&$1.57\times10^{ -2}$&$2.15\times10^{ -8}$\\
\ce{^{ 84}Se}&  34&  50&  84&0.40&1.47&$6.77\times10^{ -6}$&$5.56\times10^{ -6}$&$1.43\times10^{  2}$&$2.90\times10^{ -2}$&$1.11\times10^{ -7}$\\
\ce{^{ 82}Ge}&  32&  50&  82&0.39&1.56&$1.67\times10^{ -5}$&$1.76\times10^{ -5}$&$1.05\times10^{  2}$&$4.38\times10^{ -2}$&$3.39\times10^{ -7}$\\
\ce{^{ 80}Zn}&  30&  50&  80&0.38&1.67&$3.93\times10^{ -5}$&$5.27\times10^{ -5}$&$7.86\times10^{  1}$&$6.47\times10^{ -2}$&$9.66\times10^{ -7}$\\
\ce{^{ 78}Ni}&  28&  50&  78&0.36&1.79&$5.24\times10^{ -5}$&$7.31\times10^{ -5}$&$7.08\times10^{  1}$&$7.15\times10^{ -2}$&$1.28\times10^{ -6}$\\
\ce{^{ 80}Ni}&  28&  52&  80&0.35&1.86&$7.63\times10^{ -5}$&$1.17\times10^{ -4}$&$6.30\times10^{  1}$&$8.41\times10^{ -2}$&$2.04\times10^{ -6}$\\
\ce{^{124}Mo}&  42&  82& 124&0.34&1.95&$9.89\times10^{ -5}$&$1.56\times10^{ -4}$&$6.69\times10^{  1}$&$9.27\times10^{ -2}$&$3.62\times10^{ -6}$\\
\ce{^{122}Zr}&  40&  82& 122&0.33&2.05&$1.42\times10^{ -4}$&$2.41\times10^{ -4}$&$5.90\times10^{  1}$&$1.07\times10^{ -1}$&$5.41\times10^{ -6}$\\
\ce{^{121} Y}&  39&  82& 121&0.32&2.10&$1.65\times10^{ -4}$&$2.89\times10^{ -4}$&$5.60\times10^{  1}$&$1.14\times10^{ -1}$&$6.36\times10^{ -6}$\\
\ce{^{120}Sr}&  38&  82& 120&0.32&2.16&$1.80\times10^{ -4}$&$3.19\times10^{ -4}$&$5.41\times10^{  1}$&$1.17\times10^{ -1}$&$6.90\times10^{ -6}$\\
\ce{^{122}Sr}&  38&  84& 122&0.31&2.21&$2.26\times10^{ -4}$&$4.21\times10^{ -4}$&$5.05\times10^{  1}$&$1.29\times10^{ -1}$&$9.12\times10^{ -6}$\\
\ce{^{124}Sr}&  38&  86& 124&0.31&2.26&$2.59\times10^{ -4}$&$4.94\times10^{ -4}$&$4.85\times10^{  1}$&$1.36\times10^{ -1}$&$1.07\times10^{ -5}$\\
\hline
\hline
\end{tabular*}
\end{center}
\label{tab:HFB32-B_star=0}
\end{table*}

\begin{table*}[htbp]
\caption{
Same as Table~\ref{tab:HFB32-B_star=0}, but for $B_\star=10^3$.
}\vspace{-1mm}
\begin{center}
\begin{tabular*}{\textwidth}{@{\extracolsep{\fill}}cccccccccccc}
\hline
\hline
\multicolumn{12}{c}{AME2020\,+\,HFB32\; ($B_\star=10^3$)}\\
\hline
Nuclides&$Z$&$N$&$A$&$y_e$&$N/Z$&$n_\text{max}$\,[fm$^{-3}$]&$P_\text{max}$\,[MeV/fm$^3$]&$\nu_\text{max}$&$R$\,[fm]&$\hbar\omega$\,[MeV]&$S$\,[MeV/fm$^3$]\\
\hline
\ce{^{ 56}Fe}&  26&  30&  56&0.46&1.15&$2.62\times10^{ -6}$&$1.97\times10^{ -7}$&      0&$1.72\times10^{  2}$&$2.07\times10^{ -2}$&$3.16\times10^{ -8}$\\
\ce{^{ 62}Ni}&  28&  34&  62&0.45&1.21&$1.10\times10^{ -5}$&$6.18\times10^{ -6}$&      0&$1.10\times10^{  2}$&$4.12\times10^{ -2}$&$2.16\times10^{ -7}$\\
\ce{^{ 64}Ni}&  28&  36&  64&0.44&1.29&$1.42\times10^{ -5}$&$1.01\times10^{ -5}$&      0&$1.02\times10^{  2}$&$4.54\times10^{ -2}$&$2.93\times10^{ -7}$\\
\ce{^{ 88}Sr}&  38&  50&  88&0.43&1.32&$1.55\times10^{ -5}$&$1.16\times10^{ -5}$&      0&$1.11\times10^{  2}$&$4.68\times10^{ -2}$&$3.96\times10^{ -7}$\\
\ce{^{ 86}Kr}&  36&  50&  86&0.42&1.39&$2.60\times10^{ -5}$&$3.19\times10^{ -5}$&      0&$9.25\times10^{  1}$&$5.87\times10^{ -2}$&$7.28\times10^{ -7}$\\
\ce{^{ 84}Se}&  34&  50&  84&0.40&1.47&$3.88\times10^{ -5}$&$6.79\times10^{ -5}$&      0&$8.03\times10^{  1}$&$6.93\times10^{ -2}$&$1.14\times10^{ -6}$\\
\ce{^{ 82}Ge}&  32&  50&  82&0.39&1.56&$5.20\times10^{ -5}$&$1.15\times10^{ -4}$&      0&$7.22\times10^{  1}$&$7.75\times10^{ -2}$&$1.55\times10^{ -6}$\\
\ce{^{ 80}Zn}&  30&  50&  80&0.38&1.67&$6.98\times10^{ -5}$&$1.93\times10^{ -4}$&      0&$6.49\times10^{  1}$&$8.62\times10^{ -2}$&$2.08\times10^{ -6}$\\
\ce{^{ 78}Ni}&  28&  50&  78&0.36&1.79&$7.85\times10^{ -5}$&$2.24\times10^{ -4}$&      0&$6.19\times10^{  1}$&$8.75\times10^{ -2}$&$2.19\times10^{ -6}$\\
\ce{^{ 80}Ni}&  28&  52&  80&0.35&1.86&$8.88\times10^{ -5}$&$2.74\times10^{ -4}$&      0&$5.99\times10^{  1}$&$9.07\times10^{ -2}$&$2.50\times10^{ -6}$\\
\ce{^{124}Mo}&  42&  82& 124&0.34&1.95&$9.96\times10^{ -5}$&$3.21\times10^{ -4}$&      0&$6.67\times10^{  1}$&$9.30\times10^{ -2}$&$3.65\times10^{ -6}$\\
\ce{^{122}Zr}&  40&  82& 122&0.33&2.05&$1.15\times10^{ -4}$&$3.99\times10^{ -4}$&      0&$6.33\times10^{  1}$&$9.66\times10^{ -2}$&$4.08\times10^{ -6}$\\
\ce{^{121} Y}&  39&  82& 121&0.32&2.10&$1.21\times10^{ -4}$&$4.32\times10^{ -4}$&      0&$6.20\times10^{  1}$&$9.76\times10^{ -2}$&$4.23\times10^{ -6}$\\
\ce{^{120}Sr}&  38&  82& 120&0.32&2.16&$1.91\times10^{ -4}$&$4.68\times10^{ -4}$&  0,  1&$5.31\times10^{  1}$&$1.20\times10^{ -1}$&$7.43\times10^{ -6}$\\
\ce{^{122}Sr}&  38&  84& 122&0.31&2.21&$2.60\times10^{ -4}$&$5.83\times10^{ -4}$&      1&$4.82\times10^{  1}$&$1.38\times10^{ -1}$&$1.10\times10^{ -5}$\\
\ce{^{124}Sr}&  38&  86& 124&0.31&2.26&$2.94\times10^{ -4}$&$6.63\times10^{ -4}$&      1&$4.65\times10^{  1}$&$1.44\times10^{ -1}$&$1.26\times10^{ -5}$\\
\hline
\hline
\end{tabular*}
\end{center}
\label{tab:HFB32-B_star=1000}
\end{table*}

\begin{table*}[htbp]
\centering
\caption{
Same as Tables~\ref{tab:HFB32-B_star=0} and \ref{tab:HFB32-B_star=1000}, but for $B_\star=3\times10^4$.
}\vspace{-1mm}
\begin{center}
\begin{tabular*}{\textwidth}{@{\extracolsep{\fill}}cccccccccccc}
\hline
\hline
\multicolumn{12}{c}{AME2020\,+\,HFB32\; ($B_\star=3\times10^4$)}\\
\hline
Nuclides&$Z$&$N$&$A$&$y_e$&$N/Z$&$n_\text{max}$\,[fm$^{-3}$]&$P_\text{max}$\,[MeV/fm$^3$]&$\nu_\text{max}$&$R$\,[fm]&$\hbar\omega$\,[MeV]&$S$\,[MeV/fm$^3$]\\
\hline
\ce{^{ 60}Ni}&  28&  32&  60&0.47&1.14&$1.37\times10^{ -4}$&$1.56\times10^{ -5}$&      0&$4.71\times10^{  1}$&$1.50\times10^{ -1}$&$6.54\times10^{ -6}$\\
\ce{^{ 90}Zr}&  40&  50&  90&0.44&1.25&$2.88\times10^{ -4}$&$9.43\times10^{ -5}$&      0&$4.21\times10^{  1}$&$2.08\times10^{ -1}$&$2.09\times10^{ -5}$\\
\ce{^{ 88}Sr}&  38&  50&  88&0.43&1.32&$8.27\times10^{ -4}$&$1.02\times10^{ -3}$&      0&$2.94\times10^{  1}$&$3.42\times10^{ -1}$&$7.94\times10^{ -5}$\\
\ce{^{ 86}Kr}&  36&  50&  86&0.42&1.39&$1.01\times10^{ -3}$&$1.48\times10^{ -3}$&      0&$2.73\times10^{  1}$&$3.67\times10^{ -1}$&$9.65\times10^{ -5}$\\
\ce{^{128}Sn}&  50&  78& 128&0.39&1.56&$1.13\times10^{ -3}$&$1.54\times10^{ -3}$&      0&$3.00\times10^{  1}$&$3.61\times10^{ -1}$&$1.26\times10^{ -4}$\\
\ce{^{134}Te}&  52&  82& 134&0.39&1.58&$1.31\times10^{ -3}$&$2.10\times10^{ -3}$&      0&$2.90\times10^{  1}$&$3.87\times10^{ -1}$&$1.57\times10^{ -4}$\\
\ce{^{132}Sn}&  50&  82& 132&0.38&1.64&$2.52\times10^{ -3}$&$7.90\times10^{ -3}$&      0&$2.32\times10^{  1}$&$5.23\times10^{ -1}$&$3.53\times10^{ -4}$\\
\ce{^{128}Pd}&  46&  82& 128&0.36&1.78&$3.09\times10^{ -3}$&$1.09\times10^{ -2}$&      0&$2.15\times10^{  1}$&$5.49\times10^{ -1}$&$4.08\times10^{ -4}$\\
\ce{^{126}Ru}&  44&  82& 126&0.35&1.86&$3.23\times10^{ -3}$&$1.13\times10^{ -2}$&      0&$2.10\times10^{  1}$&$5.46\times10^{ -1}$&$4.06\times10^{ -4}$\\
\ce{^{188}Sm}&  62& 126& 188&0.33&2.03&$3.61\times10^{ -3}$&$1.24\times10^{ -2}$&      0&$2.31\times10^{  1}$&$5.45\times10^{ -1}$&$5.49\times10^{ -4}$\\
\ce{^{186}Nd}&  60& 126& 186&0.32&2.10&$3.89\times10^{ -3}$&$1.38\times10^{ -2}$&      0&$2.25\times10^{  1}$&$5.53\times10^{ -1}$&$5.75\times10^{ -4}$\\
\ce{^{184}Ce}&  58& 126& 184&0.32&2.17&$4.46\times10^{ -3}$&$1.75\times10^{ -2}$&      0&$2.14\times10^{  1}$&$5.79\times10^{ -1}$&$6.55\times10^{ -4}$\\
\ce{^{182}Ba}&  56& 126& 182&0.31&2.25&$4.73\times10^{ -3}$&$1.88\times10^{ -2}$&      0&$2.09\times10^{  1}$&$5.82\times10^{ -1}$&$6.68\times10^{ -4}$\\
\ce{^{180}Xe}&  54& 126& 180&0.30&2.33&$4.99\times10^{ -3}$&$1.99\times10^{ -2}$&      0&$2.05\times10^{  1}$&$5.82\times10^{ -1}$&$6.77\times10^{ -4}$\\
\hline
\hline
\end{tabular*}
\end{center}
\label{tab:HFB32-B_star=30000}
\end{table*}

\begin{table*}[htbp]
\caption{
Same as Tables~\ref{tab:HFB32-B_star=0}--\ref{tab:HFB32-B_star=30000}, but for $B_\star=7\times10^4$
}\vspace{-1mm}
\begin{center}
\begin{tabular*}{\textwidth}{@{\extracolsep{\fill}}cccccccccccc}
\hline
\hline
\multicolumn{12}{c}{AME2020\,+\,HFB32\; ($B_\star=7\times10^4$)}\\
\hline
Nuclides&$Z$&$N$&$A$&$y_e$&$N/Z$&$n_\text{max}$\,[fm$^{-3}$]&$P_\text{max}$\,[MeV/fm$^3$]&$\nu_\text{max}$&$R$\,[fm]&$\hbar\omega$\,[MeV]&$S$\,[MeV/fm$^3$]\\
\hline
\ce{^{ 90}Zr}&  40&  50&  90&0.44&1.25&$1.17\times10^{ -3}$&$7.47\times10^{ -4}$&      0&$2.64\times10^{  1}$&$4.18\times10^{ -1}$&$1.36\times10^{ -4}$\\
\ce{^{ 88}Sr}&  38&  50&  88&0.43&1.32&$1.95\times10^{ -3}$&$2.28\times10^{ -3}$&      0&$2.21\times10^{  1}$&$5.25\times10^{ -1}$&$2.49\times10^{ -4}$\\
\ce{^{138}Ba}&  56&  82& 138&0.41&1.46&$2.47\times10^{ -3}$&$3.10\times10^{ -3}$&      0&$2.37\times10^{  1}$&$5.55\times10^{ -1}$&$4.07\times10^{ -4}$\\
\ce{^{136}Xe}&  54&  82& 136&0.40&1.52&$3.13\times10^{ -3}$&$5.05\times10^{ -3}$&      0&$2.18\times10^{  1}$&$6.12\times10^{ -1}$&$5.30\times10^{ -4}$\\
\ce{^{134}Te}&  52&  82& 134&0.39&1.58&$3.79\times10^{ -3}$&$7.30\times10^{ -3}$&      0&$2.04\times10^{  1}$&$6.57\times10^{ -1}$&$6.45\times10^{ -4}$\\
\ce{^{132}Sn}&  50&  82& 132&0.38&1.64&$6.27\times10^{ -3}$&$2.03\times10^{ -2}$&      0&$1.71\times10^{  1}$&$8.25\times10^{ -1}$&$1.19\times10^{ -3}$\\
\ce{^{196}Yb}&  70& 126& 196&0.36&1.80&$6.99\times10^{ -3}$&$2.18\times10^{ -2}$&      0&$1.88\times10^{  1}$&$8.21\times10^{ -1}$&$1.59\times10^{ -3}$\\
\ce{^{194}Er}&  68& 126& 194&0.35&1.85&$7.45\times10^{ -3}$&$2.41\times10^{ -2}$&      0&$1.84\times10^{  1}$&$8.32\times10^{ -1}$&$1.66\times10^{ -3}$\\
\ce{^{192}Dy}&  66& 126& 192&0.34&1.91&$7.99\times10^{ -3}$&$2.69\times10^{ -2}$&      0&$1.79\times10^{  1}$&$8.45\times10^{ -1}$&$1.74\times10^{ -3}$\\
\ce{^{190}Gd}&  64& 126& 190&0.34&1.97&$8.45\times10^{ -3}$&$2.91\times10^{ -2}$&      0&$1.75\times10^{  1}$&$8.52\times10^{ -1}$&$1.79\times10^{ -3}$\\
\ce{^{276} U}&  92& 184& 276&0.33&2.00&$8.75\times10^{ -3}$&$2.94\times10^{ -2}$&      0&$1.96\times10^{  1}$&$8.58\times10^{ -1}$&$2.35\times10^{ -3}$\\
\ce{^{274}Th}&  90& 184& 274&0.33&2.04&$9.21\times10^{ -3}$&$3.18\times10^{ -2}$&      0&$1.92\times10^{  1}$&$8.67\times10^{ -1}$&$2.44\times10^{ -3}$\\
\ce{^{272}Ra}&  88& 184& 272&0.32&2.09&$9.65\times10^{ -3}$&$3.41\times10^{ -2}$&      0&$1.89\times10^{  1}$&$8.74\times10^{ -1}$&$2.50\times10^{ -3}$\\
\ce{^{270}Rn}&  86& 184& 270&0.32&2.14&$1.01\times10^{ -2}$&$3.62\times10^{ -2}$&      0&$1.86\times10^{  1}$&$8.79\times10^{ -1}$&$2.55\times10^{ -3}$\\
\ce{^{268}Po}&  84& 184& 268&0.31&2.19&$1.05\times10^{ -2}$&$3.84\times10^{ -2}$&      0&$1.82\times10^{  1}$&$8.83\times10^{ -1}$&$2.60\times10^{ -3}$\\
\ce{^{266}Pb}&  82& 184& 266&0.31&2.24&$1.23\times10^{ -2}$&$5.18\times10^{ -2}$&      0&$1.73\times10^{  1}$&$9.40\times10^{ -1}$&$3.09\times10^{ -3}$\\
\hline
\hline
\end{tabular*}
\end{center}
\label{tab:HFB32-B_star=70000}
\end{table*}

\begin{table*}[htbp]
\caption{
Same as Tables~\ref{tab:HFB32-B_star=0}--\ref{tab:HFB32-B_star=70000}, but for $B_\star=10^5$.
}\vspace{-1mm}
\begin{center}
\begin{tabular*}{\textwidth}{@{\extracolsep{\fill}}cccccccccccc}
\hline
\hline
\multicolumn{12}{c}{AME2020\,+\,HFB32\; ($B_\star=10^5$)}\\
\hline
Nuclides&$Z$&$N$&$A$&$y_e$&$N/Z$&$n_\text{max}$\,[fm$^{-3}$]&$P_\text{max}$\,[MeV/fm$^3$]&$\nu_\text{max}$&$R$\,[fm]&$\hbar\omega$\,[MeV]&$S$\,[MeV/fm$^3$]\\
\hline
\ce{^{ 90}Zr}&  40&  50&  90&0.44&1.25&$2.06\times10^{ -3}$&$1.66\times10^{ -3}$&      0&$2.18\times10^{  1}$&$5.55\times10^{ -1}$&$2.88\times10^{ -4}$\\
\ce{^{ 88}Sr}&  38&  50&  88&0.43&1.32&$2.40\times10^{ -3}$&$2.23\times10^{ -3}$&      0&$2.06\times10^{  1}$&$5.82\times10^{ -1}$&$3.28\times10^{ -4}$\\
\ce{^{118}Sn}&  50&  68& 118&0.42&1.36&$2.68\times10^{ -3}$&$2.57\times10^{ -3}$&      0&$2.19\times10^{  1}$&$6.04\times10^{ -1}$&$4.45\times10^{ -4}$\\
\ce{^{140}Ce}&  58&  82& 140&0.41&1.41&$3.12\times10^{ -3}$&$3.32\times10^{ -3}$&      0&$2.20\times10^{  1}$&$6.37\times10^{ -1}$&$5.85\times10^{ -4}$\\
\ce{^{138}Ba}&  56&  82& 138&0.41&1.46&$4.10\times10^{ -3}$&$5.98\times10^{ -3}$&      0&$2.00\times10^{  1}$&$7.15\times10^{ -1}$&$8.01\times10^{ -4}$\\
\ce{^{136}Xe}&  54&  82& 136&0.40&1.52&$5.05\times10^{ -3}$&$9.08\times10^{ -3}$&      0&$1.86\times10^{  1}$&$7.76\times10^{ -1}$&$1.00\times10^{ -3}$\\
\ce{^{134}Te}&  52&  82& 134&0.39&1.58&$5.98\times10^{ -3}$&$1.26\times10^{ -2}$&      0&$1.75\times10^{  1}$&$8.26\times10^{ -1}$&$1.19\times10^{ -3}$\\
\ce{^{132}Sn}&  50&  82& 132&0.38&1.64&$7.78\times10^{ -3}$&$2.12\times10^{ -2}$&      0&$1.59\times10^{  1}$&$9.19\times10^{ -1}$&$1.59\times10^{ -3}$\\
\ce{^{202}Os}&  76& 126& 202&0.38&1.66&$8.75\times10^{ -3}$&$2.53\times10^{ -2}$&      0&$1.77\times10^{  1}$&$9.68\times10^{ -1}$&$2.44\times10^{ -3}$\\
\ce{^{200} W}&  74& 126& 200&0.37&1.70&$9.43\times10^{ -3}$&$2.88\times10^{ -2}$&      0&$1.72\times10^{  1}$&$9.88\times10^{ -1}$&$2.58\times10^{ -3}$\\
\ce{^{198}Hf}&  72& 126& 198&0.36&1.75&$1.00\times10^{ -2}$&$3.18\times10^{ -2}$&      0&$1.68\times10^{  1}$&$1.00\times10^{  0}$&$2.69\times10^{ -3}$\\
\ce{^{266} U}&  92& 174& 266&0.35&1.89&$1.09\times10^{ -2}$&$3.28\times10^{ -2}$&      0&$1.80\times10^{  1}$&$9.92\times10^{ -1}$&$3.31\times10^{ -3}$\\
\ce{^{268} U}&  92& 176& 268&0.34&1.91&$1.11\times10^{ -2}$&$3.34\times10^{ -2}$&      0&$1.79\times10^{  1}$&$9.93\times10^{ -1}$&$3.35\times10^{ -3}$\\
\ce{^{270} U}&  92& 178& 270&0.34&1.93&$1.16\times10^{ -2}$&$3.62\times10^{ -2}$&      0&$1.77\times10^{  1}$&$1.01\times10^{  0}$&$3.51\times10^{ -3}$\\
\ce{^{272} U}&  92& 180& 272&0.34&1.96&$1.17\times10^{ -2}$&$3.69\times10^{ -2}$&      0&$1.77\times10^{  1}$&$1.01\times10^{  0}$&$3.56\times10^{ -3}$\\
\ce{^{274} U}&  92& 182& 274&0.34&1.98&$1.22\times10^{ -2}$&$3.96\times10^{ -2}$&      0&$1.75\times10^{  1}$&$1.02\times10^{  0}$&$3.71\times10^{ -3}$\\
\ce{^{276} U}&  92& 184& 276&0.33&2.00&$1.37\times10^{ -2}$&$4.98\times10^{ -2}$&      0&$1.69\times10^{  1}$&$1.07\times10^{  0}$&$4.28\times10^{ -3}$\\
\ce{^{274}Th}&  90& 184& 274&0.33&2.04&$1.43\times10^{ -2}$&$5.29\times10^{ -2}$&      0&$1.66\times10^{  1}$&$1.08\times10^{  0}$&$4.37\times10^{ -3}$\\
\ce{^{272}Ra}&  88& 184& 272&0.32&2.09&$1.49\times10^{ -2}$&$5.61\times10^{ -2}$&      0&$1.63\times10^{  1}$&$1.08\times10^{  0}$&$4.46\times10^{ -3}$\\
\ce{^{270}Rn}&  86& 184& 270&0.32&2.14&$1.55\times10^{ -2}$&$5.96\times10^{ -2}$&      0&$1.61\times10^{  1}$&$1.09\times10^{  0}$&$4.55\times10^{ -3}$\\
\ce{^{268}Po}&  84& 184& 268&0.31&2.19&$1.61\times10^{ -2}$&$6.26\times10^{ -2}$&      0&$1.58\times10^{  1}$&$1.09\times10^{  0}$&$4.61\times10^{ -3}$\\
\ce{^{266}Pb}&  82& 184& 266&0.31&2.24&$1.72\times10^{ -2}$&$6.91\times10^{ -2}$&      0&$1.55\times10^{  1}$&$1.11\times10^{  0}$&$4.82\times10^{ -3}$\\
\ce{^{375}Mc}& \textbf{115}& 260& 375&0.31&2.26&$1.79\times10^{ -2}$&$7.12\times10^{ -2}$&      0&$1.71\times10^{  1}$&$1.13\times10^{  0}$&$6.33\times10^{ -3}$\\
\ce{^{376}Mc}& \textbf{115}& 261& 376&0.31&2.27&$1.80\times10^{ -2}$&$7.20\times10^{ -2}$&      0&$1.71\times10^{  1}$&$1.13\times10^{  0}$&$6.37\times10^{ -3}$\\
\ce{^{377}Mc}& \textbf{115}& 262& 377&0.31&2.28&$1.82\times10^{ -2}$&$7.34\times10^{ -2}$&      0&$1.70\times10^{  1}$&$1.13\times10^{  0}$&$6.45\times10^{ -3}$\\
\ce{^{379}Mc}& \textbf{115}& 264& 379&0.30&2.30&$1.84\times10^{ -2}$&$7.41\times10^{ -2}$&      0&$1.70\times10^{  1}$&$1.13\times10^{  0}$&$6.49\times10^{ -3}$\\
\ce{^{396}Uue}& \textbf{119}& 277& 396&0.30&2.33&$1.88\times10^{ -2}$&$7.56\times10^{ -2}$&      0&$1.71\times10^{  1}$&$1.13\times10^{  0}$&$6.74\times10^{ -3}$\\
\ce{^{400}Uue}& \textbf{119}& 281& 400&0.30&2.36&$1.92\times10^{ -2}$&$7.79\times10^{ -2}$&      0&$1.71\times10^{  1}$&$1.13\times10^{  0}$&$6.87\times10^{ -3}$\\
\ce{^{402}Uue}& \textbf{119}& 283& 402&0.30&2.38&$1.94\times10^{ -2}$&$7.87\times10^{ -2}$&      0&$1.70\times10^{  1}$&$1.13\times10^{  0}$&$6.91\times10^{ -3}$\\
\hline
\hline
\end{tabular*}
\end{center}
\label{tab:HFB32-B_star=100000}
\end{table*}


\begin{table*}[htbp]
\caption{
Smae as Table~\ref{tab:HFB27-B_star=0} (\textit{i.e.} $B=0$),
but calculated with the BML mass model \cite{BML}.
}\vspace{-1mm}
\begin{center}
\begin{tabular*}{\textwidth}{@{\extracolsep{\fill}}ccccccccccc}
\hline
\hline
\multicolumn{11}{c}{AME2020\,+\,BML\; ($B=0$)}\\
\hline
Nuclides&$Z$&$N$&$A$&$y_e$&$N/Z$&$n_\text{max}$\,[fm$^{-3}$]&$P_\text{max}$\,[MeV/fm$^3$]&$R$\,[fm]&$\hbar\omega$\,[MeV]&$S$\,[MeV/fm$^3$]\\
\hline
\ce{^{ 56}Fe}&  26&  30&  56&0.46&1.15&$4.94\times10^{ -9}$&$3.37\times10^{-10}$&$1.39\times10^{  3}$&$8.98\times10^{ -4}$&$7.35\times10^{-12}$\\
\ce{^{ 62}Ni}&  28&  34&  62&0.45&1.21&$1.58\times10^{ -7}$&$4.16\times10^{ -8}$&$4.54\times10^{  2}$&$4.94\times10^{ -3}$&$7.55\times10^{-10}$\\
\ce{^{ 58}Fe}&  26&  32&  58&0.45&1.23&$1.65\times10^{ -7}$&$4.37\times10^{ -8}$&$4.38\times10^{  2}$&$5.01\times10^{ -3}$&$7.53\times10^{-10}$\\
\ce{^{ 64}Ni}&  28&  36&  64&0.44&1.29&$7.96\times10^{ -7}$&$3.53\times10^{ -7}$&$2.68\times10^{  2}$&$1.07\times10^{ -2}$&$6.25\times10^{ -9}$\\
\ce{^{ 66}Ni}&  28&  38&  66&0.42&1.36&$9.24\times10^{ -7}$&$4.14\times10^{ -7}$&$2.57\times10^{  2}$&$1.12\times10^{ -2}$&$7.33\times10^{ -9}$\\
\ce{^{ 86}Kr}&  36&  50&  86&0.42&1.39&$1.85\times10^{ -6}$&$1.02\times10^{ -6}$&$2.23\times10^{  2}$&$1.57\times10^{ -2}$&$2.15\times10^{ -8}$\\
\ce{^{ 84}Se}&  34&  50&  84&0.40&1.47&$6.77\times10^{ -6}$&$5.56\times10^{ -6}$&$1.43\times10^{  2}$&$2.90\times10^{ -2}$&$1.11\times10^{ -7}$\\
\ce{^{ 82}Ge}&  32&  50&  82&0.39&1.56&$1.67\times10^{ -5}$&$1.76\times10^{ -5}$&$1.05\times10^{  2}$&$4.38\times10^{ -2}$&$3.39\times10^{ -7}$\\
\ce{^{ 80}Zn}&  30&  50&  80&0.38&1.67&$3.24\times10^{ -5}$&$4.06\times10^{ -5}$&$8.39\times10^{  1}$&$5.87\times10^{ -2}$&$7.46\times10^{ -7}$\\
\ce{^{ 78}Ni}&  28&  50&  78&0.36&1.79&$7.22\times10^{ -5}$&$1.12\times10^{ -4}$&$6.36\times10^{  1}$&$8.39\times10^{ -2}$&$1.96\times10^{ -6}$\\
\ce{^{ 80}Ni}&  28&  52&  80&0.35&1.86&$8.73\times10^{ -5}$&$1.40\times10^{ -4}$&$6.03\times10^{  1}$&$9.00\times10^{ -2}$&$2.44\times10^{ -6}$\\
\ce{^{124}Mo}&  42&  82& 124&0.34&1.95&$1.07\times10^{ -4}$&$1.72\times10^{ -4}$&$6.52\times10^{  1}$&$9.62\times10^{ -2}$&$4.00\times10^{ -6}$\\
\ce{^{122}Zr}&  40&  82& 122&0.33&2.05&$1.56\times10^{ -4}$&$2.75\times10^{ -4}$&$5.71\times10^{  1}$&$1.13\times10^{ -1}$&$6.16\times10^{ -6}$\\
\ce{^{120}Sr}&  38&  82& 120&0.32&2.16&$2.24\times10^{ -4}$&$4.25\times10^{ -4}$&$5.04\times10^{  1}$&$1.30\times10^{ -1}$&$9.21\times10^{ -6}$\\
\ce{^{118}Kr}&  36&  82& 118&0.31&2.28&$2.58\times10^{ -4}$&$4.89\times10^{ -4}$&$4.78\times10^{  1}$&$1.35\times10^{ -1}$&$1.02\times10^{ -5}$\\
\hline
\hline
\end{tabular*}
\end{center}
\label{tab:BML-B_star=0}
\end{table*}

\begin{table*}[htbp]
\caption{
Smae as Table~\ref{tab:BML-B_star=0}, but for $B_\star=10^3$.
}\vspace{-1mm}
\begin{center}
\begin{tabular*}{\textwidth}{@{\extracolsep{\fill}}cccccccccccc}
\hline
\hline
\multicolumn{12}{c}{AME2020\,+\,BML\; ($B_\star=10^3$)}\\
\hline
Nuclides&$Z$&$N$&$A$&$y_e$&$N/Z$&$n_\text{max}$\,[fm$^{-3}$]&$P_\text{max}$\,[MeV/fm$^3$]&$\nu_\text{max}$&$R$\,[fm]&$\hbar\omega$\,[MeV]&$S$\,[MeV/fm$^3$]\\
\hline
\ce{^{ 56}Fe}&  26&  30&  56&0.46&1.15&$2.62\times10^{ -6}$&$1.97\times10^{ -7}$&      0&$1.72\times10^{  2}$&$2.07\times10^{ -2}$&$3.16\times10^{ -8}$\\
\ce{^{ 62}Ni}&  28&  34&  62&0.45&1.21&$1.10\times10^{ -5}$&$6.18\times10^{ -6}$&      0&$1.10\times10^{  2}$&$4.12\times10^{ -2}$&$2.16\times10^{ -7}$\\
\ce{^{ 64}Ni}&  28&  36&  64&0.44&1.29&$1.42\times10^{ -5}$&$1.01\times10^{ -5}$&      0&$1.02\times10^{  2}$&$4.54\times10^{ -2}$&$2.93\times10^{ -7}$\\
\ce{^{ 88}Sr}&  38&  50&  88&0.43&1.32&$1.55\times10^{ -5}$&$1.16\times10^{ -5}$&      0&$1.11\times10^{  2}$&$4.68\times10^{ -2}$&$3.96\times10^{ -7}$\\
\ce{^{ 86}Kr}&  36&  50&  86&0.42&1.39&$2.60\times10^{ -5}$&$3.19\times10^{ -5}$&      0&$9.25\times10^{  1}$&$5.87\times10^{ -2}$&$7.28\times10^{ -7}$\\
\ce{^{ 84}Se}&  34&  50&  84&0.40&1.47&$3.88\times10^{ -5}$&$6.79\times10^{ -5}$&      0&$8.03\times10^{  1}$&$6.93\times10^{ -2}$&$1.14\times10^{ -6}$\\
\ce{^{ 82}Ge}&  32&  50&  82&0.39&1.56&$5.20\times10^{ -5}$&$1.15\times10^{ -4}$&      0&$7.22\times10^{  1}$&$7.75\times10^{ -2}$&$1.55\times10^{ -6}$\\
\ce{^{ 80}Zn}&  30&  50&  80&0.38&1.67&$6.55\times10^{ -5}$&$1.70\times10^{ -4}$&      0&$6.63\times10^{  1}$&$8.35\times10^{ -2}$&$1.91\times10^{ -6}$\\
\ce{^{ 78}Ni}&  28&  50&  78&0.36&1.79&$8.70\times10^{ -5}$&$2.76\times10^{ -4}$&      0&$5.98\times10^{  1}$&$9.21\times10^{ -2}$&$2.51\times10^{ -6}$\\
\ce{^{ 80}Ni}&  28&  52&  80&0.35&1.86&$9.33\times10^{ -5}$&$3.02\times10^{ -4}$&      0&$5.89\times10^{  1}$&$9.30\times10^{ -2}$&$2.67\times10^{ -6}$\\
\ce{^{124}Mo}&  42&  82& 124&0.34&1.95&$1.02\times10^{ -4}$&$3.37\times10^{ -4}$&      0&$6.62\times10^{  1}$&$9.41\times10^{ -2}$&$3.77\times10^{ -6}$\\
\ce{^{122}Zr}&  40&  82& 122&0.33&2.05&$1.18\times10^{ -4}$&$4.24\times10^{ -4}$&      0&$6.27\times10^{  1}$&$9.80\times10^{ -2}$&$4.25\times10^{ -6}$\\
\ce{^{120}Sr}&  38&  82& 120&0.32&2.16&$2.58\times10^{ -4}$&$5.89\times10^{ -4}$&  0,  1&$4.81\times10^{  1}$&$1.40\times10^{ -1}$&$1.11\times10^{ -5}$\\
\ce{^{118}Kr}&  36&  82& 118&0.31&2.28&$2.92\times10^{ -4}$&$6.57\times10^{ -4}$&      1&$4.58\times10^{  1}$&$1.43\times10^{ -1}$&$1.20\times10^{ -5}$\\
\hline
\hline
\end{tabular*}
\end{center}
\label{tab:BML-B_star=1000}
\end{table*}

\begin{table*}[htbp]
\caption{
Smae as Tables~\ref{tab:BML-B_star=0} and \ref{tab:BML-B_star=1000}, but for $B_\star=3\times10^4$.
}\vspace{-1mm}
\begin{center}
\begin{tabular*}{\textwidth}{@{\extracolsep{\fill}}cccccccccccc}
\hline
\hline
\multicolumn{12}{c}{AME2020\,+\,BML\; ($B_\star=3\times10^4$)}\\
\hline
Nuclides&$Z$&$N$&$A$&$y_e$&$N/Z$&$n_\text{max}$\,[fm$^{-3}$]&$P_\text{max}$\,[MeV/fm$^3$]&$\nu_\text{max}$&$R$\,[fm]&$\hbar\omega$\,[MeV]&$S$\,[MeV/fm$^3$]\\
\hline
\ce{^{ 60}Ni}&  28&  32&  60&0.47&1.14&$1.37\times10^{ -4}$&$1.56\times10^{ -5}$&      0&$4.71\times10^{  1}$&$1.50\times10^{ -1}$&$6.54\times10^{ -6}$\\
\ce{^{ 90}Zr}&  40&  50&  90&0.44&1.25&$2.88\times10^{ -4}$&$9.43\times10^{ -5}$&      0&$4.21\times10^{  1}$&$2.08\times10^{ -1}$&$2.09\times10^{ -5}$\\
\ce{^{ 88}Sr}&  38&  50&  88&0.43&1.32&$8.27\times10^{ -4}$&$1.02\times10^{ -3}$&      0&$2.94\times10^{  1}$&$3.42\times10^{ -1}$&$7.94\times10^{ -5}$\\
\ce{^{ 86}Kr}&  36&  50&  86&0.42&1.39&$1.01\times10^{ -3}$&$1.48\times10^{ -3}$&      0&$2.73\times10^{  1}$&$3.67\times10^{ -1}$&$9.65\times10^{ -5}$\\
\ce{^{128}Sn}&  50&  78& 128&0.39&1.56&$1.13\times10^{ -3}$&$1.54\times10^{ -3}$&      0&$3.00\times10^{  1}$&$3.61\times10^{ -1}$&$1.26\times10^{ -4}$\\
\ce{^{134}Te}&  52&  82& 134&0.39&1.58&$1.31\times10^{ -3}$&$2.10\times10^{ -3}$&      0&$2.90\times10^{  1}$&$3.87\times10^{ -1}$&$1.57\times10^{ -4}$\\
\ce{^{132}Sn}&  50&  82& 132&0.38&1.64&$2.53\times10^{ -3}$&$7.98\times10^{ -3}$&      0&$2.32\times10^{  1}$&$5.24\times10^{ -1}$&$3.55\times10^{ -4}$\\
\ce{^{130}Cd}&  48&  82& 130&0.37&1.71&$2.75\times10^{ -3}$&$8.99\times10^{ -3}$&      0&$2.24\times10^{  1}$&$5.32\times10^{ -1}$&$3.73\times10^{ -4}$\\
\ce{^{128}Pd}&  46&  82& 128&0.36&1.78&$3.04\times10^{ -3}$&$1.05\times10^{ -2}$&      0&$2.16\times10^{  1}$&$5.45\times10^{ -1}$&$4.01\times10^{ -4}$\\
\ce{^{126}Ru}&  44&  82& 126&0.35&1.86&$3.16\times10^{ -3}$&$1.07\times10^{ -2}$&      0&$2.12\times10^{  1}$&$5.40\times10^{ -1}$&$3.93\times10^{ -4}$\\
\ce{^{166}Ba}&  56& 110& 166&0.34&1.96&$3.39\times10^{ -3}$&$1.14\times10^{ -2}$&      0&$2.27\times10^{  1}$&$5.40\times10^{ -1}$&$4.85\times10^{ -4}$\\
\ce{^{168}Ba}&  56& 112& 168&0.33&2.00&$3.58\times10^{ -3}$&$1.25\times10^{ -2}$&      0&$2.24\times10^{  1}$&$5.49\times10^{ -1}$&$5.14\times10^{ -4}$\\
\ce{^{170}Ba}&  56& 114& 170&0.33&2.04&$3.73\times10^{ -3}$&$1.33\times10^{ -2}$&      0&$2.21\times10^{  1}$&$5.53\times10^{ -1}$&$5.34\times10^{ -4}$\\
\ce{^{172}Ba}&  56& 116& 172&0.33&2.07&$4.00\times10^{ -3}$&$1.49\times10^{ -2}$&      0&$2.17\times10^{  1}$&$5.66\times10^{ -1}$&$5.77\times10^{ -4}$\\
\ce{^{182}Ce}&  58& 124& 182&0.32&2.14&$4.17\times10^{ -3}$&$1.55\times10^{ -2}$&      0&$2.18\times10^{  1}$&$5.66\times10^{ -1}$&$6.07\times10^{ -4}$\\
\ce{^{184}Ce}&  58& 126& 184&0.32&2.17&$4.42\times10^{ -3}$&$1.72\times10^{ -2}$&      0&$2.15\times10^{  1}$&$5.76\times10^{ -1}$&$6.47\times10^{ -4}$\\
\ce{^{182}Ba}&  56& 126& 182&0.31&2.25&$4.77\times10^{ -3}$&$1.92\times10^{ -2}$&      0&$2.09\times10^{  1}$&$5.84\times10^{ -1}$&$6.77\times10^{ -4}$\\
\ce{^{180}Xe}&  54& 126& 180&0.30&2.33&$4.99\times10^{ -3}$&$1.99\times10^{ -2}$&      0&$2.05\times10^{  1}$&$5.82\times10^{ -1}$&$6.77\times10^{ -4}$\\
\hline
\hline
\end{tabular*}
\end{center}
\label{tab:BML-B_star=30000}
\end{table*}

\begin{table*}[htbp]
\caption{
Smae as Tables~\ref{tab:BML-B_star=0}--\ref{tab:BML-B_star=30000}, but for $B_\star=7\times10^4$.
}\vspace{-1mm}
\begin{center}
\begin{tabular*}{\textwidth}{@{\extracolsep{\fill}}cccccccccccc}
\hline
\hline
\multicolumn{12}{c}{AME2020\,+\,BML\; ($B_\star=7\times10^4$)}\\
\hline
Nuclides&$Z$&$N$&$A$&$y_e$&$N/Z$&$n_\text{max}$\,[fm$^{-3}$]&$P_\text{max}$\,[MeV/fm$^3$]&$\nu_\text{max}$&$R$\,[fm]&$\hbar\omega$\,[MeV]&$S$\,[MeV/fm$^3$]\\
\hline
\ce{^{ 90}Zr}&  40&  50&  90&0.44&1.25&$1.17\times10^{ -3}$&$7.47\times10^{ -4}$&      0&$2.64\times10^{  1}$&$4.18\times10^{ -1}$&$1.36\times10^{ -4}$\\
\ce{^{ 88}Sr}&  38&  50&  88&0.43&1.32&$1.95\times10^{ -3}$&$2.28\times10^{ -3}$&      0&$2.21\times10^{  1}$&$5.25\times10^{ -1}$&$2.49\times10^{ -4}$\\
\ce{^{138}Ba}&  56&  82& 138&0.41&1.46&$2.47\times10^{ -3}$&$3.10\times10^{ -3}$&      0&$2.37\times10^{  1}$&$5.55\times10^{ -1}$&$4.07\times10^{ -4}$\\
\ce{^{136}Xe}&  54&  82& 136&0.40&1.52&$3.13\times10^{ -3}$&$5.05\times10^{ -3}$&      0&$2.18\times10^{  1}$&$6.12\times10^{ -1}$&$5.30\times10^{ -4}$\\
\ce{^{134}Te}&  52&  82& 134&0.39&1.58&$3.79\times10^{ -3}$&$7.30\times10^{ -3}$&      0&$2.04\times10^{  1}$&$6.57\times10^{ -1}$&$6.45\times10^{ -4}$\\
\ce{^{132}Sn}&  50&  82& 132&0.38&1.64&$6.30\times10^{ -3}$&$2.05\times10^{ -2}$&      0&$1.71\times10^{  1}$&$8.27\times10^{ -1}$&$1.20\times10^{ -3}$\\
\ce{^{208} W}&  74& 134& 208&0.36&1.81&$7.07\times10^{ -3}$&$2.20\times10^{ -2}$&      0&$1.91\times10^{  1}$&$8.23\times10^{ -1}$&$1.67\times10^{ -3}$\\
\ce{^{204}Hf}&  72& 132& 204&0.35&1.83&$7.18\times10^{ -3}$&$2.25\times10^{ -2}$&      0&$1.89\times10^{  1}$&$8.23\times10^{ -1}$&$1.66\times10^{ -3}$\\
\ce{^{206}Hf}&  72& 134& 206&0.35&1.86&$7.85\times10^{ -3}$&$2.66\times10^{ -2}$&      0&$1.84\times10^{  1}$&$8.52\times10^{ -1}$&$1.84\times10^{ -3}$\\
\ce{^{204}Yb}&  70& 134& 204&0.34&1.91&$8.30\times10^{ -3}$&$2.88\times10^{ -2}$&      0&$1.80\times10^{  1}$&$8.60\times10^{ -1}$&$1.90\times10^{ -3}$\\
\ce{^{276} U}&  92& 184& 276&0.33&2.00&$8.92\times10^{ -3}$&$3.06\times10^{ -2}$&      0&$1.95\times10^{  1}$&$8.66\times10^{ -1}$&$2.42\times10^{ -3}$\\
\ce{^{272}Th}&  90& 182& 272&0.33&2.02&$9.02\times10^{ -3}$&$3.09\times10^{ -2}$&      0&$1.93\times10^{  1}$&$8.64\times10^{ -1}$&$2.39\times10^{ -3}$\\
\ce{^{274}Th}&  90& 184& 274&0.33&2.04&$9.56\times10^{ -3}$&$3.44\times10^{ -2}$&      0&$1.90\times10^{  1}$&$8.83\times10^{ -1}$&$2.56\times10^{ -3}$\\
\ce{^{272}Ra}&  88& 184& 272&0.32&2.09&$1.02\times10^{ -2}$&$3.81\times10^{ -2}$&      0&$1.85\times10^{  1}$&$8.96\times10^{ -1}$&$2.68\times10^{ -3}$\\
\ce{^{270}Rn}&  86& 184& 270&0.32&2.14&$1.07\times10^{ -2}$&$4.12\times10^{ -2}$&      0&$1.82\times10^{  1}$&$9.06\times10^{ -1}$&$2.77\times10^{ -3}$\\
\ce{^{268}Po}&  84& 184& 268&0.31&2.19&$1.11\times10^{ -2}$&$4.33\times10^{ -2}$&      0&$1.79\times10^{  1}$&$9.09\times10^{ -1}$&$2.81\times10^{ -3}$\\
\ce{^{266}Pb}&  82& 184& 266&0.31&2.24&$1.22\times10^{ -2}$&$5.13\times10^{ -2}$&      0&$1.73\times10^{  1}$&$9.37\times10^{ -1}$&$3.07\times10^{ -3}$\\
\hline
\hline
\end{tabular*}
\end{center}
\label{tab:BML-B_star=70000}
\end{table*}

\begin{table*}[htbp]
\caption{
Smae as Tables~\ref{tab:BML-B_star=0}--\ref{tab:BML-B_star=70000}, but for $B_\star=10^5$.
}\vspace{-1mm}
\begin{center}
\begin{tabular*}{\textwidth}{@{\extracolsep{\fill}}cccccccccccc}
\hline
\hline
\multicolumn{12}{c}{AME2020\,+\,BML\; ($B_\star=10^5$)}\\
\hline
Nuclides&$Z$&$N$&$A$&$y_e$&$N/Z$&$n_\text{max}$\,[fm$^{-3}$]&$P_\text{max}$\,[MeV/fm$^3$]&$\nu_\text{max}$&$R$\,[fm]&$\hbar\omega$\,[MeV]&$S$\,[MeV/fm$^3$]\\
\hline
\ce{^{ 90}Zr}&  40&  50&  90&0.44&1.25&$2.06\times10^{ -3}$&$1.66\times10^{ -3}$&      0&$2.18\times10^{  1}$&$5.55\times10^{ -1}$&$2.88\times10^{ -4}$\\
\ce{^{ 88}Sr}&  38&  50&  88&0.43&1.32&$2.40\times10^{ -3}$&$2.23\times10^{ -3}$&      0&$2.06\times10^{  1}$&$5.82\times10^{ -1}$&$3.28\times10^{ -4}$\\
\ce{^{118}Sn}&  50&  68& 118&0.42&1.36&$2.68\times10^{ -3}$&$2.57\times10^{ -3}$&      0&$2.19\times10^{  1}$&$6.04\times10^{ -1}$&$4.45\times10^{ -4}$\\
\ce{^{140}Ce}&  58&  82& 140&0.41&1.41&$3.12\times10^{ -3}$&$3.32\times10^{ -3}$&      0&$2.20\times10^{  1}$&$6.37\times10^{ -1}$&$5.85\times10^{ -4}$\\
\ce{^{138}Ba}&  56&  82& 138&0.41&1.46&$4.10\times10^{ -3}$&$5.98\times10^{ -3}$&      0&$2.00\times10^{  1}$&$7.15\times10^{ -1}$&$8.01\times10^{ -4}$\\
\ce{^{136}Xe}&  54&  82& 136&0.40&1.52&$5.05\times10^{ -3}$&$9.08\times10^{ -3}$&      0&$1.86\times10^{  1}$&$7.76\times10^{ -1}$&$1.00\times10^{ -3}$\\
\ce{^{134}Te}&  52&  82& 134&0.39&1.58&$5.98\times10^{ -3}$&$1.26\times10^{ -2}$&      0&$1.75\times10^{  1}$&$8.26\times10^{ -1}$&$1.19\times10^{ -3}$\\
\ce{^{132}Sn}&  50&  82& 132&0.38&1.64&$7.56\times10^{ -3}$&$1.99\times10^{ -2}$&      0&$1.61\times10^{  1}$&$9.06\times10^{ -1}$&$1.53\times10^{ -3}$\\
\ce{^{204}Pt}&  78& 126& 204&0.38&1.62&$7.95\times10^{ -3}$&$2.12\times10^{ -2}$&      0&$1.83\times10^{  1}$&$9.38\times10^{ -1}$&$2.23\times10^{ -3}$\\
\ce{^{202}Os}&  76& 126& 202&0.38&1.66&$8.91\times10^{ -3}$&$2.63\times10^{ -2}$&      0&$1.75\times10^{  1}$&$9.77\times10^{ -1}$&$2.50\times10^{ -3}$\\
\ce{^{200} W}&  74& 126& 200&0.37&1.70&$9.47\times10^{ -3}$&$2.91\times10^{ -2}$&      0&$1.71\times10^{  1}$&$9.91\times10^{ -1}$&$2.60\times10^{ -3}$\\
\ce{^{208} W}&  74& 134& 208&0.36&1.81&$1.05\times10^{ -2}$&$3.34\times10^{ -2}$&      0&$1.68\times10^{  1}$&$1.00\times10^{  0}$&$2.84\times10^{ -3}$\\
\ce{^{272}Pu}&  94& 178& 272&0.35&1.89&$1.11\times10^{ -2}$&$3.41\times10^{ -2}$&      0&$1.80\times10^{  1}$&$1.00\times10^{  0}$&$3.44\times10^{ -3}$\\
\ce{^{268} U}&  92& 176& 268&0.34&1.91&$1.14\times10^{ -2}$&$3.55\times10^{ -2}$&      0&$1.78\times10^{  1}$&$1.01\times10^{  0}$&$3.47\times10^{ -3}$\\
\ce{^{270} U}&  92& 178& 270&0.34&1.93&$1.17\times10^{ -2}$&$3.73\times10^{ -2}$&      0&$1.76\times10^{  1}$&$1.01\times10^{  0}$&$3.58\times10^{ -3}$\\
\ce{^{272} U}&  92& 180& 272&0.34&1.96&$1.22\times10^{ -2}$&$4.00\times10^{ -2}$&      0&$1.75\times10^{  1}$&$1.03\times10^{  0}$&$3.74\times10^{ -3}$\\
\ce{^{274} U}&  92& 182& 274&0.34&1.98&$1.27\times10^{ -2}$&$4.33\times10^{ -2}$&      0&$1.72\times10^{  1}$&$1.04\times10^{  0}$&$3.93\times10^{ -3}$\\
\ce{^{276} U}&  92& 184& 276&0.33&2.00&$1.40\times10^{ -2}$&$5.18\times10^{ -2}$&      0&$1.68\times10^{  1}$&$1.08\times10^{  0}$&$4.39\times10^{ -3}$\\
\ce{^{274}Th}&  90& 184& 274&0.33&2.04&$1.45\times10^{ -2}$&$5.45\times10^{ -2}$&      0&$1.65\times10^{  1}$&$1.09\times10^{  0}$&$4.45\times10^{ -3}$\\
\ce{^{338}Ds}& \textbf{110}& 228& 338&0.33&2.07&$1.52\times10^{ -2}$&$5.78\times10^{ -2}$&      0&$1.74\times10^{  1}$&$1.10\times10^{  0}$&$5.38\times10^{ -3}$\\
\ce{^{339}Mt}& \textbf{109}& 230& 339&0.32&2.11&$1.56\times10^{ -2}$&$5.96\times10^{ -2}$&      0&$1.73\times10^{  1}$&$1.10\times10^{  0}$&$5.44\times10^{ -3}$\\
\ce{^{338}Hs}& \textbf{108}& 230& 338&0.32&2.13&$1.63\times10^{ -2}$&$6.45\times10^{ -2}$&      0&$1.70\times10^{  1}$&$1.12\times10^{  0}$&$5.67\times10^{ -3}$\\
\ce{^{339}Bh}& \textbf{107}& 232& 339&0.32&2.17&$1.68\times10^{ -2}$&$6.71\times10^{ -2}$&      0&$1.69\times10^{  1}$&$1.12\times10^{  0}$&$5.78\times10^{ -3}$\\
\ce{^{338}Sg}& \textbf{106}& 232& 338&0.31&2.19&$1.74\times10^{ -2}$&$7.20\times10^{ -2}$&      0&$1.67\times10^{  1}$&$1.14\times10^{  0}$&$5.99\times10^{ -3}$\\
\ce{^{339}Db}& \textbf{105}& 234& 339&0.31&2.23&$1.81\times10^{ -2}$&$7.56\times10^{ -2}$&      0&$1.65\times10^{  1}$&$1.14\times10^{  0}$&$6.13\times10^{ -3}$\\
\ce{^{338}Rf}& \textbf{104}& 234& 338&0.31&2.25&$1.84\times10^{ -2}$&$7.79\times10^{ -2}$&      0&$1.64\times10^{  1}$&$1.15\times10^{  0}$&$6.20\times10^{ -3}$\\
\hline
\hline
\end{tabular*}
\end{center}
\label{tab:BML-B_star=100000}
\end{table*}


\begin{table*}[htbp]
\caption{
Smae as Table~\ref{tab:HFB27-B_star=0} (\textit{i.e.} $B=0$),
but calculated with the KTUY mass model \cite{KTUY}.
}\vspace{-1mm}
\begin{center}
\begin{tabular*}{\textwidth}{@{\extracolsep{\fill}}ccccccccccc}
\hline
\hline
\multicolumn{11}{c}{AME2020\,+\,KTUY\; ($B=0$)}\\
\hline
Nuclides&$Z$&$N$&$A$&$y_e$&$N/Z$&$n_\text{max}$\,[fm$^{-3}$]&$P_\text{max}$\,[MeV/fm$^3$]&$R$\,[fm]&$\hbar\omega$\,[MeV]&$S$\,[MeV/fm$^3$]\\
\hline
\ce{^{ 56}Fe}&  26&  30&  56&0.46&1.15&$4.94\times10^{ -9}$&$3.37\times10^{-10}$&$1.39\times10^{  3}$&$8.98\times10^{ -4}$&$7.35\times10^{-12}$\\
\ce{^{ 62}Ni}&  28&  34&  62&0.45&1.21&$1.58\times10^{ -7}$&$4.16\times10^{ -8}$&$4.54\times10^{  2}$&$4.94\times10^{ -3}$&$7.55\times10^{-10}$\\
\ce{^{ 58}Fe}&  26&  32&  58&0.45&1.23&$1.65\times10^{ -7}$&$4.37\times10^{ -8}$&$4.38\times10^{  2}$&$5.01\times10^{ -3}$&$7.53\times10^{-10}$\\
\ce{^{ 64}Ni}&  28&  36&  64&0.44&1.29&$7.96\times10^{ -7}$&$3.53\times10^{ -7}$&$2.68\times10^{  2}$&$1.07\times10^{ -2}$&$6.25\times10^{ -9}$\\
\ce{^{ 66}Ni}&  28&  38&  66&0.42&1.36&$9.24\times10^{ -7}$&$4.14\times10^{ -7}$&$2.57\times10^{  2}$&$1.12\times10^{ -2}$&$7.33\times10^{ -9}$\\
\ce{^{ 86}Kr}&  36&  50&  86&0.42&1.39&$1.85\times10^{ -6}$&$1.02\times10^{ -6}$&$2.23\times10^{  2}$&$1.57\times10^{ -2}$&$2.15\times10^{ -8}$\\
\ce{^{ 84}Se}&  34&  50&  84&0.40&1.47&$6.77\times10^{ -6}$&$5.56\times10^{ -6}$&$1.43\times10^{  2}$&$2.90\times10^{ -2}$&$1.11\times10^{ -7}$\\
\ce{^{ 82}Ge}&  32&  50&  82&0.39&1.56&$1.67\times10^{ -5}$&$1.76\times10^{ -5}$&$1.05\times10^{  2}$&$4.38\times10^{ -2}$&$3.39\times10^{ -7}$\\
\ce{^{ 80}Zn}&  30&  50&  80&0.38&1.67&$2.15\times10^{ -5}$&$2.35\times10^{ -5}$&$9.62\times10^{  1}$&$4.78\times10^{ -2}$&$4.32\times10^{ -7}$\\
\ce{^{ 78}Ni}&  28&  50&  78&0.36&1.79&$8.14\times10^{ -5}$&$1.31\times10^{ -4}$&$6.12\times10^{  1}$&$8.91\times10^{ -2}$&$2.30\times10^{ -6}$\\
\ce{^{124}Mo}&  42&  82& 124&0.34&1.95&$1.01\times10^{ -4}$&$1.60\times10^{ -4}$&$6.64\times10^{  1}$&$9.37\times10^{ -2}$&$3.73\times10^{ -6}$\\
\ce{^{122}Zr}&  40&  82& 122&0.33&2.05&$1.47\times10^{ -4}$&$2.54\times10^{ -4}$&$5.83\times10^{  1}$&$1.09\times10^{ -1}$&$5.69\times10^{ -6}$\\
\ce{^{120}Sr}&  38&  82& 120&0.32&2.16&$2.29\times10^{ -4}$&$4.38\times10^{ -4}$&$5.00\times10^{  1}$&$1.32\times10^{ -1}$&$9.48\times10^{ -6}$\\
\ce{^{122}Sr}&  38&  84& 122&0.31&2.21&$2.35\times10^{ -4}$&$4.43\times10^{ -4}$&$4.99\times10^{  1}$&$1.31\times10^{ -1}$&$9.58\times10^{ -6}$\\
\ce{^{119}Kr}&  36&  83& 119&0.30&2.31&$2.47\times10^{ -4}$&$4.56\times10^{ -4}$&$4.86\times10^{  1}$&$1.31\times10^{ -1}$&$9.50\times10^{ -6}$\\
\ce{^{120}Kr}&  36&  84& 120&0.30&2.33&$2.64\times10^{ -4}$&$4.94\times10^{ -4}$&$4.77\times10^{  1}$&$1.34\times10^{ -1}$&$1.03\times10^{ -5}$\\
\hline
\hline
\end{tabular*}
\end{center}
\label{tab:KTUY-B_star=0}
\end{table*}

\begin{table*}[htbp]
\caption{
Smae as Table~\ref{tab:KTUY-B_star=0}, but for $B_\star=10^3$.
}\vspace{-1mm}
\begin{center}
\begin{tabular*}{\textwidth}{@{\extracolsep{\fill}}cccccccccccc}
\hline
\hline
\multicolumn{12}{c}{AME2020\,+\,KTUY\; ($B_\star=10^3$)}\\
\hline
Nuclides&$Z$&$N$&$A$&$y_e$&$N/Z$&$n_\text{max}$\,[fm$^{-3}$]&$P_\text{max}$\,[MeV/fm$^3$]&$\nu_\text{max}$&$R$\,[fm]&$\hbar\omega$\,[MeV]&$S$\,[MeV/fm$^3$]\\
\hline
\ce{^{ 56}Fe}&  26&  30&  56&0.46&1.15&$2.62\times10^{ -6}$&$1.97\times10^{ -7}$&      0&$1.72\times10^{  2}$&$2.07\times10^{ -2}$&$3.16\times10^{ -8}$\\
\ce{^{ 62}Ni}&  28&  34&  62&0.45&1.21&$1.10\times10^{ -5}$&$6.18\times10^{ -6}$&      0&$1.10\times10^{  2}$&$4.12\times10^{ -2}$&$2.16\times10^{ -7}$\\
\ce{^{ 64}Ni}&  28&  36&  64&0.44&1.29&$1.42\times10^{ -5}$&$1.01\times10^{ -5}$&      0&$1.02\times10^{  2}$&$4.54\times10^{ -2}$&$2.93\times10^{ -7}$\\
\ce{^{ 88}Sr}&  38&  50&  88&0.43&1.32&$1.55\times10^{ -5}$&$1.16\times10^{ -5}$&      0&$1.11\times10^{  2}$&$4.68\times10^{ -2}$&$3.96\times10^{ -7}$\\
\ce{^{ 86}Kr}&  36&  50&  86&0.42&1.39&$2.60\times10^{ -5}$&$3.19\times10^{ -5}$&      0&$9.25\times10^{  1}$&$5.87\times10^{ -2}$&$7.28\times10^{ -7}$\\
\ce{^{ 84}Se}&  34&  50&  84&0.40&1.47&$3.88\times10^{ -5}$&$6.79\times10^{ -5}$&      0&$8.03\times10^{  1}$&$6.93\times10^{ -2}$&$1.14\times10^{ -6}$\\
\ce{^{ 82}Ge}&  32&  50&  82&0.39&1.56&$5.20\times10^{ -5}$&$1.15\times10^{ -4}$&      0&$7.22\times10^{  1}$&$7.75\times10^{ -2}$&$1.55\times10^{ -6}$\\
\ce{^{ 80}Zn}&  30&  50&  80&0.38&1.67&$5.79\times10^{ -5}$&$1.32\times10^{ -4}$&      0&$6.91\times10^{  1}$&$7.85\times10^{ -2}$&$1.62\times10^{ -6}$\\
\ce{^{ 78}Ni}&  28&  50&  78&0.36&1.79&$8.96\times10^{ -5}$&$2.93\times10^{ -4}$&      0&$5.92\times10^{  1}$&$9.35\times10^{ -2}$&$2.62\times10^{ -6}$\\
\ce{^{124}Mo}&  42&  82& 124&0.34&1.95&$1.01\times10^{ -4}$&$3.27\times10^{ -4}$&      0&$6.65\times10^{  1}$&$9.35\times10^{ -2}$&$3.70\times10^{ -6}$\\
\ce{^{122}Zr}&  40&  82& 122&0.33&2.05&$1.16\times10^{ -4}$&$4.07\times10^{ -4}$&      0&$6.31\times10^{  1}$&$9.70\times10^{ -2}$&$4.14\times10^{ -6}$\\
\ce{^{120}Sr}&  38&  82& 120&0.32&2.16&$2.62\times10^{ -4}$&$6.00\times10^{ -4}$&  0,  1&$4.78\times10^{  1}$&$1.41\times10^{ -1}$&$1.14\times10^{ -5}$\\
\ce{^{122}Sr}&  38&  84& 122&0.31&2.21&$2.71\times10^{ -4}$&$6.12\times10^{ -4}$&      1&$4.75\times10^{  1}$&$1.41\times10^{ -1}$&$1.16\times10^{ -5}$\\
\ce{^{119}Kr}&  36&  83& 119&0.30&2.31&$2.81\times10^{ -4}$&$6.19\times10^{ -4}$&      1&$4.66\times10^{  1}$&$1.40\times10^{ -1}$&$1.13\times10^{ -5}$\\
\ce{^{120}Kr}&  36&  84& 120&0.30&2.33&$3.00\times10^{ -4}$&$6.63\times10^{ -4}$&      1&$4.57\times10^{  1}$&$1.43\times10^{ -1}$&$1.22\times10^{ -5}$\\
\hline
\hline
\end{tabular*}
\end{center}
\label{tab:KTUY-B_star=1000}
\end{table*}

\begin{table*}[htbp]
\caption{
Smae as Tables~\ref{tab:KTUY-B_star=0} and \ref{tab:KTUY-B_star=1000}, but for $B_\star=3\times10^4$.
}\vspace{-1mm}
\begin{center}
\begin{tabular*}{\textwidth}{@{\extracolsep{\fill}}cccccccccccc}
\hline
\hline
\multicolumn{12}{c}{AME2020\,+\,KTUY\; ($B_\star=3\times10^4$)}\\
\hline
Nuclides&$Z$&$N$&$A$&$y_e$&$N/Z$&$n_\text{max}$\,[fm$^{-3}$]&$P_\text{max}$\,[MeV/fm$^3$]&$\nu_\text{max}$&$R$\,[fm]&$\hbar\omega$\,[MeV]&$S$\,[MeV/fm$^3$]\\
\hline
\ce{^{ 60}Ni}&  28&  32&  60&0.47&1.14&$1.37\times10^{ -4}$&$1.56\times10^{ -5}$&      0&$4.71\times10^{  1}$&$1.50\times10^{ -1}$&$6.54\times10^{ -6}$\\
\ce{^{ 90}Zr}&  40&  50&  90&0.44&1.25&$2.88\times10^{ -4}$&$9.43\times10^{ -5}$&      0&$4.21\times10^{  1}$&$2.08\times10^{ -1}$&$2.09\times10^{ -5}$\\
\ce{^{ 88}Sr}&  38&  50&  88&0.43&1.32&$8.27\times10^{ -4}$&$1.02\times10^{ -3}$&      0&$2.94\times10^{  1}$&$3.42\times10^{ -1}$&$7.94\times10^{ -5}$\\
\ce{^{ 86}Kr}&  36&  50&  86&0.42&1.39&$1.01\times10^{ -3}$&$1.48\times10^{ -3}$&      0&$2.73\times10^{  1}$&$3.67\times10^{ -1}$&$9.65\times10^{ -5}$\\
\ce{^{128}Sn}&  50&  78& 128&0.39&1.56&$1.13\times10^{ -3}$&$1.54\times10^{ -3}$&      0&$3.00\times10^{  1}$&$3.61\times10^{ -1}$&$1.26\times10^{ -4}$\\
\ce{^{134}Te}&  52&  82& 134&0.39&1.58&$1.31\times10^{ -3}$&$2.10\times10^{ -3}$&      0&$2.90\times10^{  1}$&$3.87\times10^{ -1}$&$1.57\times10^{ -4}$\\
\ce{^{132}Sn}&  50&  82& 132&0.38&1.64&$2.48\times10^{ -3}$&$7.67\times10^{ -3}$&      0&$2.33\times10^{  1}$&$5.19\times10^{ -1}$&$3.46\times10^{ -4}$\\
\ce{^{128}Pd}&  46&  82& 128&0.36&1.78&$2.97\times10^{ -3}$&$1.00\times10^{ -2}$&      0&$2.17\times10^{  1}$&$5.39\times10^{ -1}$&$3.88\times10^{ -4}$\\
\ce{^{126}Ru}&  44&  82& 126&0.35&1.86&$3.20\times10^{ -3}$&$1.11\times10^{ -2}$&      0&$2.11\times10^{  1}$&$5.44\times10^{ -1}$&$4.01\times10^{ -4}$\\
\ce{^{188}Sm}&  62& 126& 188&0.33&2.03&$3.60\times10^{ -3}$&$1.22\times10^{ -2}$&      0&$2.32\times10^{  1}$&$5.44\times10^{ -1}$&$5.45\times10^{ -4}$\\
\ce{^{186}Nd}&  60& 126& 186&0.32&2.10&$3.91\times10^{ -3}$&$1.39\times10^{ -2}$&      0&$2.25\times10^{  1}$&$5.55\times10^{ -1}$&$5.79\times10^{ -4}$\\
\ce{^{184}Ce}&  58& 126& 184&0.32&2.17&$4.60\times10^{ -3}$&$1.86\times10^{ -2}$&      0&$2.12\times10^{  1}$&$5.88\times10^{ -1}$&$6.81\times10^{ -4}$\\
\ce{^{182}Ba}&  56& 126& 182&0.31&2.25&$4.82\times10^{ -3}$&$1.95\times10^{ -2}$&      0&$2.08\times10^{  1}$&$5.87\times10^{ -1}$&$6.86\times10^{ -4}$\\
\ce{^{180}Xe}&  54& 126& 180&0.30&2.33&$4.99\times10^{ -3}$&$1.99\times10^{ -2}$&      0&$2.05\times10^{  1}$&$5.82\times10^{ -1}$&$6.77\times10^{ -4}$\\
\hline
\hline
\end{tabular*}
\end{center}
\label{tab:KTUY-B_star=30000}
\end{table*}

\begin{table*}[htbp]
\caption{
Smae as Tables~\ref{tab:KTUY-B_star=0}--\ref{tab:KTUY-B_star=30000}, but for $B_\star=7\times10^4$.
}\vspace{-1mm}
\begin{center}
\begin{tabular*}{\textwidth}{@{\extracolsep{\fill}}cccccccccccc}
\hline
\hline
\multicolumn{12}{c}{AME2020\,+\,KTUY\; ($B_\star=7\times10^4$)}\\
\hline
Nuclides&$Z$&$N$&$A$&$y_e$&$N/Z$&$n_\text{max}$\,[fm$^{-3}$]&$P_\text{max}$\,[MeV/fm$^3$]&$\nu_\text{max}$&$R$\,[fm]&$\hbar\omega$\,[MeV]&$S$\,[MeV/fm$^3$]\\
\hline
\ce{^{ 90}Zr}&  40&  50&  90&0.44&1.25&$1.17\times10^{ -3}$&$7.47\times10^{ -4}$&      0&$2.64\times10^{  1}$&$4.18\times10^{ -1}$&$1.36\times10^{ -4}$\\
\ce{^{ 88}Sr}&  38&  50&  88&0.43&1.32&$1.95\times10^{ -3}$&$2.28\times10^{ -3}$&      0&$2.21\times10^{  1}$&$5.25\times10^{ -1}$&$2.49\times10^{ -4}$\\
\ce{^{138}Ba}&  56&  82& 138&0.41&1.46&$2.47\times10^{ -3}$&$3.10\times10^{ -3}$&      0&$2.37\times10^{  1}$&$5.55\times10^{ -1}$&$4.07\times10^{ -4}$\\
\ce{^{136}Xe}&  54&  82& 136&0.40&1.52&$3.13\times10^{ -3}$&$5.05\times10^{ -3}$&      0&$2.18\times10^{  1}$&$6.12\times10^{ -1}$&$5.30\times10^{ -4}$\\
\ce{^{134}Te}&  52&  82& 134&0.39&1.58&$3.79\times10^{ -3}$&$7.30\times10^{ -3}$&      0&$2.04\times10^{  1}$&$6.57\times10^{ -1}$&$6.45\times10^{ -4}$\\
\ce{^{132}Sn}&  50&  82& 132&0.38&1.64&$6.03\times10^{ -3}$&$1.88\times10^{ -2}$&      0&$1.73\times10^{  1}$&$8.10\times10^{ -1}$&$1.13\times10^{ -3}$\\
\ce{^{196}Yb}&  70& 126& 196&0.36&1.80&$6.96\times10^{ -3}$&$2.16\times10^{ -2}$&      0&$1.89\times10^{  1}$&$8.19\times10^{ -1}$&$1.58\times10^{ -3}$\\
\ce{^{194}Er}&  68& 126& 194&0.35&1.85&$7.56\times10^{ -3}$&$2.48\times10^{ -2}$&      0&$1.83\times10^{  1}$&$8.38\times10^{ -1}$&$1.69\times10^{ -3}$\\
\ce{^{192}Dy}&  66& 126& 192&0.34&1.91&$8.10\times10^{ -3}$&$2.77\times10^{ -2}$&      0&$1.78\times10^{  1}$&$8.51\times10^{ -1}$&$1.77\times10^{ -3}$\\
\ce{^{190}Gd}&  64& 126& 190&0.34&1.97&$8.49\times10^{ -3}$&$2.94\times10^{ -2}$&      0&$1.75\times10^{  1}$&$8.54\times10^{ -1}$&$1.80\times10^{ -3}$\\
\ce{^{248}Pb}&  82& 166& 248&0.33&2.02&$8.89\times10^{ -3}$&$3.03\times10^{ -2}$&      0&$1.88\times10^{  1}$&$8.57\times10^{ -1}$&$2.20\times10^{ -3}$\\
\ce{^{250}Pb}&  82& 168& 250&0.33&2.05&$9.09\times10^{ -3}$&$3.12\times10^{ -2}$&      0&$1.87\times10^{  1}$&$8.60\times10^{ -1}$&$2.24\times10^{ -3}$\\
\ce{^{252}Pb}&  82& 170& 252&0.33&2.07&$9.42\times10^{ -3}$&$3.31\times10^{ -2}$&      0&$1.85\times10^{  1}$&$8.68\times10^{ -1}$&$2.33\times10^{ -3}$\\
\ce{^{254}Pb}&  82& 172& 254&0.32&2.10&$1.00\times10^{ -2}$&$3.73\times10^{ -2}$&      0&$1.82\times10^{  1}$&$8.90\times10^{ -1}$&$2.51\times10^{ -3}$\\
\ce{^{256}Pb}&  82& 174& 256&0.32&2.12&$1.03\times10^{ -2}$&$3.88\times10^{ -2}$&      0&$1.81\times10^{  1}$&$8.95\times10^{ -1}$&$2.58\times10^{ -3}$\\
\ce{^{258}Pb}&  82& 176& 258&0.32&2.15&$1.08\times10^{ -2}$&$4.25\times10^{ -2}$&      0&$1.78\times10^{  1}$&$9.10\times10^{ -1}$&$2.73\times10^{ -3}$\\
\ce{^{260}Pb}&  82& 178& 260&0.32&2.17&$1.11\times10^{ -2}$&$4.38\times10^{ -2}$&      0&$1.77\times10^{  1}$&$9.13\times10^{ -1}$&$2.78\times10^{ -3}$\\
\ce{^{262}Pb}&  82& 180& 262&0.31&2.20&$1.12\times10^{ -2}$&$4.42\times10^{ -2}$&      0&$1.77\times10^{  1}$&$9.12\times10^{ -1}$&$2.80\times10^{ -3}$\\
\ce{^{264}Pb}&  82& 182& 264&0.31&2.22&$1.16\times10^{ -2}$&$4.69\times10^{ -2}$&      0&$1.75\times10^{  1}$&$9.21\times10^{ -1}$&$2.90\times10^{ -3}$\\
\ce{^{266}Pb}&  82& 184& 266&0.31&2.24&$1.22\times10^{ -2}$&$5.13\times10^{ -2}$&      0&$1.73\times10^{  1}$&$9.37\times10^{ -1}$&$3.07\times10^{ -3}$\\
\hline
\hline
\end{tabular*}
\end{center}
\label{tab:KTUY-B_star=70000}
\end{table*}

\clearpage
\begin{table*}[htbp]
\caption{
Smae as Tables~\ref{tab:KTUY-B_star=0}--\ref{tab:KTUY-B_star=70000}, but for $B_\star=10^5$.
}\vspace{-1mm}
\begin{center}
\begin{tabular*}{\textwidth}{@{\extracolsep{\fill}}cccccccccccc}
\hline
\hline
\multicolumn{12}{c}{AME2020\,+\,KTUY\; ($B_\star=10^5$)}\\
\hline
Nuclides&$Z$&$N$&$A$&$y_e$&$N/Z$&$n_\text{max}$\,[fm$^{-3}$]&$P_\text{max}$\,[MeV/fm$^3$]&$\nu_\text{max}$&$R$\,[fm]&$\hbar\omega$\,[MeV]&$S$\,[MeV/fm$^3$]\\
\hline
\ce{^{ 90}Zr}&  40&  50&  90&0.44&1.25&$2.06\times10^{ -3}$&$1.66\times10^{ -3}$&      0&$2.18\times10^{  1}$&$5.55\times10^{ -1}$&$2.88\times10^{ -4}$\\
\ce{^{ 88}Sr}&  38&  50&  88&0.43&1.32&$2.40\times10^{ -3}$&$2.23\times10^{ -3}$&      0&$2.06\times10^{  1}$&$5.82\times10^{ -1}$&$3.28\times10^{ -4}$\\
\ce{^{118}Sn}&  50&  68& 118&0.42&1.36&$2.68\times10^{ -3}$&$2.57\times10^{ -3}$&      0&$2.19\times10^{  1}$&$6.04\times10^{ -1}$&$4.45\times10^{ -4}$\\
\ce{^{140}Ce}&  58&  82& 140&0.41&1.41&$3.12\times10^{ -3}$&$3.32\times10^{ -3}$&      0&$2.20\times10^{  1}$&$6.37\times10^{ -1}$&$5.85\times10^{ -4}$\\
\ce{^{138}Ba}&  56&  82& 138&0.41&1.46&$4.10\times10^{ -3}$&$5.98\times10^{ -3}$&      0&$2.00\times10^{  1}$&$7.15\times10^{ -1}$&$8.01\times10^{ -4}$\\
\ce{^{136}Xe}&  54&  82& 136&0.40&1.52&$5.05\times10^{ -3}$&$9.08\times10^{ -3}$&      0&$1.86\times10^{  1}$&$7.76\times10^{ -1}$&$1.00\times10^{ -3}$\\
\ce{^{134}Te}&  52&  82& 134&0.39&1.58&$5.98\times10^{ -3}$&$1.26\times10^{ -2}$&      0&$1.75\times10^{  1}$&$8.26\times10^{ -1}$&$1.19\times10^{ -3}$\\
\ce{^{132}Sn}&  50&  82& 132&0.38&1.64&$7.29\times10^{ -3}$&$1.84\times10^{ -2}$&      0&$1.63\times10^{  1}$&$8.90\times10^{ -1}$&$1.46\times10^{ -3}$\\
\ce{^{202}Os}&  76& 126& 202&0.38&1.66&$8.44\times10^{ -3}$&$2.34\times10^{ -2}$&      0&$1.79\times10^{  1}$&$9.51\times10^{ -1}$&$2.32\times10^{ -3}$\\
\ce{^{200} W}&  74& 126& 200&0.37&1.70&$9.21\times10^{ -3}$&$2.74\times10^{ -2}$&      0&$1.73\times10^{  1}$&$9.77\times10^{ -1}$&$2.51\times10^{ -3}$\\
\ce{^{198}Hf}&  72& 126& 198&0.36&1.75&$9.99\times10^{ -3}$&$3.15\times10^{ -2}$&      0&$1.68\times10^{  1}$&$9.99\times10^{ -1}$&$2.68\times10^{ -3}$\\
\ce{^{196}Yb}&  70& 126& 196&0.36&1.80&$1.05\times10^{ -2}$&$3.38\times10^{ -2}$&      0&$1.65\times10^{  1}$&$1.01\times10^{  0}$&$2.74\times10^{ -3}$\\
\ce{^{260}Th}&  90& 170& 260&0.35&1.89&$1.11\times10^{ -2}$&$3.48\times10^{ -2}$&      0&$1.77\times10^{  1}$&$1.01\times10^{  0}$&$3.37\times10^{ -3}$\\
\ce{^{256}Ra}&  88& 168& 256&0.34&1.91&$1.15\times10^{ -2}$&$3.66\times10^{ -2}$&      0&$1.75\times10^{  1}$&$1.01\times10^{  0}$&$3.42\times10^{ -3}$\\
\ce{^{258}Ra}&  88& 170& 258&0.34&1.93&$1.17\times10^{ -2}$&$3.77\times10^{ -2}$&      0&$1.74\times10^{  1}$&$1.02\times10^{  0}$&$3.48\times10^{ -3}$\\
\ce{^{272} U}&  92& 180& 272&0.34&1.96&$1.24\times10^{ -2}$&$4.12\times10^{ -2}$&      0&$1.74\times10^{  1}$&$1.03\times10^{  0}$&$3.81\times10^{ -3}$\\
\ce{^{274} U}&  92& 182& 274&0.34&1.98&$1.27\times10^{ -2}$&$4.29\times10^{ -2}$&      0&$1.73\times10^{  1}$&$1.04\times10^{  0}$&$3.90\times10^{ -3}$\\
\ce{^{276} U}&  92& 184& 276&0.33&2.00&$1.36\times10^{ -2}$&$4.93\times10^{ -2}$&      0&$1.69\times10^{  1}$&$1.07\times10^{  0}$&$4.25\times10^{ -3}$\\
\ce{^{274}Th}&  90& 184& 274&0.33&2.04&$1.43\times10^{ -2}$&$5.34\times10^{ -2}$&      0&$1.66\times10^{  1}$&$1.08\times10^{  0}$&$4.40\times10^{ -3}$\\
\ce{^{272}Ra}&  88& 184& 272&0.32&2.09&$1.51\times10^{ -2}$&$5.78\times10^{ -2}$&      0&$1.63\times10^{  1}$&$1.09\times10^{  0}$&$4.54\times10^{ -3}$\\
\ce{^{270}Rn}&  86& 184& 270&0.32&2.14&$1.63\times10^{ -2}$&$6.64\times10^{ -2}$&      0&$1.58\times10^{  1}$&$1.12\times10^{  0}$&$4.87\times10^{ -3}$\\
\ce{^{268}Po}&  84& 184& 268&0.31&2.19&$1.69\times10^{ -2}$&$6.91\times10^{ -2}$&      0&$1.56\times10^{  1}$&$1.12\times10^{  0}$&$4.91\times10^{ -3}$\\
\ce{^{264}Pb}&  82& 182& 264&0.31&2.22&$1.73\times10^{ -2}$&$7.12\times10^{ -2}$&      0&$1.54\times10^{  1}$&$1.12\times10^{  0}$&$4.91\times10^{ -3}$\\
\ce{^{266}Pb}&  82& 184& 266&0.31&2.24&$1.82\times10^{ -2}$&$7.79\times10^{ -2}$&      0&$1.52\times10^{  1}$&$1.14\times10^{  0}$&$5.20\times10^{ -3}$\\
\hline
\hline
\end{tabular*}
\end{center}
\label{tab:KTUY-B_star=100000}
\end{table*}

\begin{table*}[htbp]
\caption{
Smae as Tables~\ref{tab:KTUY-B_star=0}--\ref{tab:KTUY-B_star=100000}, but for $B_\star=1.4\times10^5$.
}\vspace{-1mm}
\begin{center}
\begin{tabular*}{\textwidth}{@{\extracolsep{\fill}}cccccccccccc}
\hline
\hline
\multicolumn{12}{c}{AME2020\,+\,KTUY\; ($B_\star=1.4\times10^5$)}\\
\hline
Nuclides&$Z$&$N$&$A$&$y_e$&$N/Z$&$n_\text{max}$\,[fm$^{-3}$]&$P_\text{max}$\,[MeV/fm$^3$]&$\nu_\text{max}$&$R$\,[fm]&$\hbar\omega$\,[MeV]&$S$\,[MeV/fm$^3$]\\
\hline
\ce{^{ 90}Zr}&  40&  50&  90&0.44&1.25&$2.78\times10^{ -3}$&$2.02\times10^{ -3}$&      0&$1.98\times10^{  1}$&$6.45\times10^{ -1}$&$4.31\times10^{ -4}$\\
\ce{^{116}Sn}&  50&  66& 116&0.43&1.32&$3.29\times10^{ -3}$&$2.59\times10^{ -3}$&      0&$2.03\times10^{  1}$&$6.80\times10^{ -1}$&$6.00\times10^{ -4}$\\
\ce{^{142}Nd}&  60&  82& 142&0.42&1.37&$3.89\times10^{ -3}$&$3.46\times10^{ -3}$&      0&$2.06\times10^{  1}$&$7.25\times10^{ -1}$&$8.24\times10^{ -4}$\\
\ce{^{140}Ce}&  58&  82& 140&0.41&1.41&$5.30\times10^{ -3}$&$6.94\times10^{ -3}$&      0&$1.85\times10^{  1}$&$8.30\times10^{ -1}$&$1.19\times10^{ -3}$\\
\ce{^{138}Ba}&  56&  82& 138&0.41&1.46&$6.39\times10^{ -3}$&$1.02\times10^{ -2}$&      0&$1.73\times10^{  1}$&$8.93\times10^{ -1}$&$1.44\times10^{ -3}$\\
\ce{^{208}Pb}&  82& 126& 208&0.39&1.54&$9.58\times10^{ -3}$&$2.18\times10^{ -2}$&      0&$1.73\times10^{  1}$&$1.06\times10^{  0}$&$3.08\times10^{ -3}$\\
\ce{^{206}Hg}&  80& 126& 206&0.39&1.57&$1.06\times10^{ -2}$&$2.63\times10^{ -2}$&      0&$1.67\times10^{  1}$&$1.10\times10^{  0}$&$3.38\times10^{ -3}$\\
\ce{^{204}Pt}&  78& 126& 204&0.38&1.62&$1.17\times10^{ -2}$&$3.18\times10^{ -2}$&      0&$1.61\times10^{  1}$&$1.14\times10^{  0}$&$3.71\times10^{ -3}$\\
\ce{^{202}Os}&  76& 126& 202&0.38&1.66&$1.29\times10^{ -2}$&$3.84\times10^{ -2}$&      0&$1.55\times10^{  1}$&$1.17\times10^{  0}$&$4.08\times10^{ -3}$\\
\ce{^{288}Rf}& \textbf{104}& 184& 288&0.36&1.77&$1.41\times10^{ -2}$&$4.04\times10^{ -2}$&      0&$1.69\times10^{  1}$&$1.18\times10^{  0}$&$5.38\times10^{ -3}$\\
\ce{^{284}No}& 102& 182& 284&0.36&1.78&$1.44\times10^{ -2}$&$4.16\times10^{ -2}$&      0&$1.68\times10^{  1}$&$1.18\times10^{  0}$&$5.39\times10^{ -3}$\\
\ce{^{286}No}& 102& 184& 286&0.36&1.80&$1.50\times10^{ -2}$&$4.51\times10^{ -2}$&      0&$1.66\times10^{  1}$&$1.20\times10^{  0}$&$5.66\times10^{ -3}$\\
\ce{^{282}Fm}& 100& 182& 282&0.35&1.82&$1.52\times10^{ -2}$&$4.60\times10^{ -2}$&      0&$1.64\times10^{  1}$&$1.20\times10^{  0}$&$5.64\times10^{ -3}$\\
\ce{^{284}Fm}& 100& 184& 284&0.35&1.84&$1.54\times10^{ -2}$&$4.69\times10^{ -2}$&      0&$1.64\times10^{  1}$&$1.20\times10^{  0}$&$5.71\times10^{ -3}$\\
\ce{^{308}Hs}& \textbf{108}& 200& 308&0.35&1.85&$1.64\times10^{ -2}$&$5.23\times10^{ -2}$&      0&$1.65\times10^{  1}$&$1.23\times10^{  0}$&$6.48\times10^{ -3}$\\
\ce{^{306}Sg}& \textbf{106}& 200& 306&0.35&1.89&$1.75\times10^{ -2}$&$5.90\times10^{ -2}$&      0&$1.61\times10^{  1}$&$1.26\times10^{  0}$&$6.86\times10^{ -3}$\\
\ce{^{304}Rf}& \textbf{104}& 200& 304&0.34&1.92&$1.86\times10^{ -2}$&$6.58\times10^{ -2}$&      0&$1.57\times10^{  1}$&$1.28\times10^{  0}$&$7.23\times10^{ -3}$\\
\ce{^{302}No}& 102& 200& 302&0.34&1.96&$1.96\times10^{ -2}$&$7.20\times10^{ -2}$&      0&$1.54\times10^{  1}$&$1.30\times10^{  0}$&$7.52\times10^{ -3}$\\
\ce{^{300}Fm}& 100& 200& 300&0.33&2.00&$2.08\times10^{ -2}$&$8.03\times10^{ -2}$&      0&$1.51\times10^{  1}$&$1.32\times10^{  0}$&$7.92\times10^{ -3}$\\
\ce{^{298}Cf}&  98& 200& 298&0.33&2.04&$2.20\times10^{ -2}$&$8.78\times10^{ -2}$&      0&$1.48\times10^{  1}$&$1.34\times10^{  0}$&$8.24\times10^{ -3}$\\
\ce{^{296}Cm}&  96& 200& 296&0.32&2.08&$2.32\times10^{ -2}$&$9.60\times10^{ -2}$&      0&$1.45\times10^{  1}$&$1.36\times10^{  0}$&$8.58\times10^{ -3}$\\
\ce{^{294}Pu}&  94& 200& 294&0.32&2.13&$2.43\times10^{ -2}$&$1.03\times10^{ -1}$&      0&$1.42\times10^{  1}$&$1.37\times10^{  0}$&$8.81\times10^{ -3}$\\
\ce{^{292} U}&  92& 200& 292&0.32&2.17&$2.54\times10^{ -2}$&$1.10\times10^{ -1}$&      0&$1.40\times10^{  1}$&$1.38\times10^{  0}$&$9.05\times10^{ -3}$\\
\ce{^{290}Th}&  90& 200& 290&0.31&2.22&$2.64\times10^{ -2}$&$1.16\times10^{ -1}$&      0&$1.38\times10^{  1}$&$1.39\times10^{  0}$&$9.19\times10^{ -3}$\\
\hline
\hline
\end{tabular*}
\end{center}
\label{tab:KTUY-B_star=140000}
\end{table*}


\begin{thebibliography}{99}

\bibitem{Hofmann(2000)}
S.~Hofmann and G.~M\"unzenberg, The discovery of the heaviest elements,
Rev. Mod. Phys. \textbf{72}, 733 (2000).
\bibitem{Armbruster(2000)}
P.~Armbruster, On the Production of Superheavy Elements,
Ann. Rev. Part. Sci. \textbf{50}, 411 (2000).
\bibitem{Hamilton(2013)}
J.H.~Hamilton, S.~Hofmann, and Y.T.~Oganessian, Search for Superheavy Nuclei,
Ann. Rev. Part. Sci. \textbf{63}, 383 (2013).
\bibitem{Oganessian(2017)}
Y.T.~Oganessian, A.~Sobiczewski, and G.M. Ter-Akopian, Superheavy nuclei: from predictions to discovery,
Phys. Scr. \textbf{92}, 023003 (2017).
\bibitem{Giuliani(2019)}
Colloquium: Superheavy elements: Oganesson and beyond,
S.A.~Giuliani, Z.~Matheson, W.~Nazarewicz, E.~Olsen, P.-G.~Reinhard, J.~Sadhukhan, B.~Schuetrumpf, N.~Schunck, and P.~Schwerdtfeger,
Rev. Mod. Phys. \textbf{91}, 011001 (2019).


\bibitem{HFB27}
S.~Goriely, N.~Chamel, and J.M.~Pearson, Hartree-Fock-Bogoliubov nuclear mass model with 0.50 MeV accuracy based on standard forms of Skyrme and pairing functionals, Phys. Rev. C \textbf{88}, 061302(R) (2013).
\bibitem{AME2020-I}
W.J.~Huang, M.~Wang, F.G.~Kondev, G.~Audi, and S.~Naimi,
The AME 2020 atomic mass evaluation (I). Evaluation of input data, and adjustment procedures,
Chinese Phys. C \textbf{45}, 030002 (2021).
\bibitem{AME2020-II}
M.~Wang, W.J.~Huang, F.G.~Kondev, G.~Audi, and S.~Naimi,
The AME 2020 atomic mass evaluation (II). Tables, graphs and references,
Chinese Phys. C \textbf{45}, 030003 (2021).


\bibitem{NeutronStars1}
P.~Haensel, A.Y.~Potekhin, and D.G.~Yakovlev, \textit{Neutron Stars 1: Equation of State and Structure}
(Springer, New York, 2007).
\bibitem{Seiradakis(2004)}
J.H.~Seiradakis and R.~Wielebinski, Morphology and characteristics of radio pulsars,
Astron. Astrophys. Rev. \textbf{12}, 239 (2004).
\bibitem{Ng(2011)}
C.-Y.~Ng and V.M.~Kaspi, High Magnetic Field Rotation ]powered Pulsars,
AIP Conf. Proc. \textbf{1379}, 60 (2011).
\bibitem{Mereghetti(2008)}
S.~Mereghetti, The strongest cosmic magnets: soft gamma-ray repeaters and anomalous X-ray pulsars,
Astron. Astrophys. Rev. \textbf{15}, 225 (2008).
\bibitem{Olausen(2014)}
S.A.~Olausen and V.M.~Kaspi, THE McGILL MAGNETAR CATALOG,
Astrophys. J., Suppl. Ser. \textbf{212}, 6 (2014).
\bibitem{Tiengo(2013)}
A.~Tiengo, P.~Esposito, S.~Mereghetti, R.~Turolla, L.~Nobili, F.~Gastaldello,
D.~G\"otz, G.L.~Israel, N.~Rea, L.~Stella, S.~Zane, and G.F.~Bignami,
A variable absorption feature in the X-ray spectrum of a magnetar,
Nature (London) \textbf{500}, 312 (2013).


\bibitem{Potekhin(1996)}
A.Y.~Potekhin, and D.G.~Yakovlev,
Electron conduction along quantizing magnetic fields in neutron star crusts II. Practical formulae,
Astron. Astrophys. \textbf{314}, 341 (1996); Erratum: \textit{ibid.} \textbf{327}, 442 (1997).
\bibitem{Potekhin(1999)}
A.Y.~Potekhin, Electron conduction in magnetized neutron star envelopes,
Astron. Astrophys. \textbf{351}, 787 (1999).
\bibitem{Broderick(2000)}
A.~Broderick, M.~Prakash, and J.M.~Lattimer,
The Equation of State of Neutron Star Matter in Strong Magnetic Fields,
Astrophys. J. \textbf{537}, 351 (2000).
\bibitem{Ventura(2001)}
J.~Ventura and A.Y.~Potekhin, Neutron Star Envelopes and
Thermal Radiation from the Magnetic Surface, The Neutron
Star-Black Hole Connection, in Proceedings of the NATO
Advanced Study Institute, Elounda, Crete, Greece, June 7-18,
1999, NATO Science Series C, Vol. \textbf{567}, edited by
C. Kouveliotou, J.E.~Ventura, and E.P.~van den Heuvel
(Springer, Netherlands, 2001), pp. 393-414.


\bibitem{Fushiki(1989)}
I.~Fushiki, E.H.~Gudmundsson, and C.J.~Pethick, Surface structure of neutron stars with high magnetic fields,
Astrophys. J. \textbf{342}, 958 (1989).

\bibitem{Lai(1991)}
D.~Lai and S.L.~Shapiro,
Cold equation of state in a strong magnetic field: Effects of inverse $\beta$-decay,
Astrophys. J. \textbf{383}, 745 (1991).

\bibitem{Kondratyev(2000)}
V.N.~Kondratyev, T.~Maruyama, and S.~Chiba, Shell Structure of Nuclei in Strong Magnetic Fields in Neutron Star Crusts,
Phys. Rev. Lett. \textbf{84}, 1086 (2000).
\bibitem{Kondratyev(2001)}
V.N.~Kondratyev, T.~Maruyama, and S.~Chiba,
Astrophys. J. \textbf{546}, 1137 (2001).

\bibitem{Arteaga(2011)}
D.~Pe\~na Arteaga, M.~Grasso, E.~Khan, and P.~Ring,
Nuclear structure in strong magnetic fields: Nuclei in the crust of a magnetar,
Phys. Rev. C \textbf{84}, 045806 (2011).
\bibitem{Basilico(2015)}
D.~Basilico, D.P.~Arteaga, X.~Roca-Maza, and G.~Col\`o, Outer crust of a cold non-accreting magnetar,
Phys. Rev. C \textbf{92}, 035802 (2015).

\bibitem{Chamel(2012)}
N.~Chamel, R.L.~Pavlov, L.M.~Mihailov, Ch.J.~Velchev, Zh.K.~Stoyanov, Y.D.~Mutafchieva,
M.D.~Ivanovich, J.M.~Pearson, and S.~Goriely,
Properties of the outer crust of strongly magnetized neutron stars from Hartree-Fock-Bogoliubov atomic mass models,
Phys. Rev. C \textbf{86}, 055804 (2012).
\bibitem{Chamel(2020)2}
N.~Chamel and Zh.K.~Stoyanov, Analytical determination of the structure of the outer crust of a cold nonaccreted neutron star: Extension to strongly quantizing magnetic fields,
Phys. Rev. C \textbf{101}, 065802 (2020).

\bibitem{Parmar(2022)}
V.~Parmar, H.C.~Das, M.K.~Sharma, and S.K.~Patra,
Magnetized neutron star crust within effective relativistic mean-field model,
arXiv:2211.07339v1 [astro-ph.HE].

\bibitem{Rabi(1928)}
I.I.~Rabi, Das freie Elektron im homogenen Magnetfeld nach der Diracschen Theorie, Z. Physik \textbf{49}, 507 (1928).
\bibitem{Landau(1930)}
L.~Landau, Diamagnetismus der Metalle, Z. Physik \textbf{64}, 629 (1930).

\bibitem{Stein(2016)_matter}
M.~Stein, A.~Sedrakian, X.-G.~Huang, and J.W.~Clark,
Spin-polarized neutron matter: Critical unpairing and BCS-BEC precursor,
Phys. Rev. C \textbf{93}, 015802 (2016).
\bibitem{Stein(2016)_C-O-Ne}
M.~Stein, J.~Maruhn, and A.~Sedrakian, Carbon-oxygen-neon mass nuclei in superstrong magnetic fields,
Phys. Rev. C \textbf{94}, 035802 (2016).

\bibitem{HFB30-32}
S.~Goriely, N.~Chamel, and J.M.~Pearson,
Further explorations of Skyrme-Hartree-Fock-Bogoliubov mass formulas. XVI. Inclusion of self-energy effects in pairing,
Phys. Rev. C \textbf{93}, 034337 (2016).
\bibitem{BML}
Z.M.~Niu and H.Z.~Liang, Nuclear mass predictions with machine learning reaching the accuracy required by r-process studies,
Phys. Rev. C \textbf{106}, L021303 (2022).
\bibitem{KTUY}
H.~Koura, T.~Tachibana, M.~Uno, and M.~Yamada, Nuclidic Mass Formula on a Spherical Basis with an Improved Even-Odd Term,
Prog. Theor. Phys. \textbf{113}, 305 (2005).

\bibitem{Iida(2020)}
K.~Iida and T.~Fujie, On the Stability of Giant Nuclei in Supernova Matter with Respect to Deconfinement,
JPS Conf. Proc. \textbf{31}, 011057 (2020).

\bibitem{Aggarwal(2021)}
S.~Aggarwal, B.~Mukhopadhyay, and G.~Gregori,
Relativistic Landau quantization in non-uniform magnetic field and its applications to white dwarfs and quantum information,
SciPost Phys. \textbf{11}, 093 (2021).

\bibitem{BPS(1971)}
G.~Baym, C.~Pethick, and P.~Sutherland, The ground state of matter at high densities:
Equation of state and stellar models, Astrophs. J. \textbf{170}, 299 (1971).

\bibitem{Baiko(2009)}
D.A.~Baiko, Coulomb crystals in the magnetic field,
Phys. Rev. E \textbf{80}, 046405 (2009).

\bibitem{Lunney(2003)}
D.~Lunney, J.M.~Pearson, and C.~Thibault, Recent trends in the determination of nuclear masses,
Rev. Mod. Phys. \textbf{75}, 1021 (2003).

\bibitem{Baiko(2001)}
D.A.~Baiko, A.Y.~Potekhin, and D.G.~Yakovlev, Thermodynamic functions of harmonic Coulomb crystals,
Phys. Rev. E \textbf{64}, 057402 (2001).

\bibitem{VanVleck(1932)}
J.H.~Van~Vleck, \textit{The Theory of Electric and Magnetic Susceptibilities}
(Oxford University Press, London, 1932).

\bibitem{SM}
See Supplmental Material at \{URL will be provided by the publisher\} for
complete lists of the outer crust compositions and complementally figures.

\bibitem{Ogata(1990)}
S.~Ogata and S.~Ichimaru, First-principles calculations of shear moduli for Monte Carlo-simulated Coulomb solids,
Phys. Rev. A \textbf{42}, 4867 (1990).

\bibitem{Fantina(2020)}
A.F.~Fantina, S.~De~Ridder, N.~Chamel, and F.~Gulminelli,
Crystallization of the outer crust of a non-accreting neutron star,
A\&A \textbf{633}, A149 (2020).

\bibitem{S_of_Chamel(2012)}
We note that the effective shear modulus $S$ in Fig.~5 of Ref.~\cite{Chamel(2012)}
contains some error. We have confirmed that their results agree with ours with
$B_\star$\,$=$\,0 and 1.4\,$\times$\,10$^3$ for $P$\,$\le$\,8\,$\times$\,10$^{-4}$\,MeV\,fm$^{-3}$,
after some correction. (N.~Chamel, Private communications.)

\bibitem{Horowitz(2009)}
C.J.~Horowitz and K.~Kadau, Breaking Strain of Neutron Star Crust and Gravitational Waves,
Phys. Rev. Lett. \textbf{102}, 191102 (2009).

\bibitem{Agbemava(2021)} S.E.~Agbemava and A.V.~Afanasjev, Hyperheavy spherical and toroidal nuclei: The role of shell structure,
Phys. Rev. C \textbf{103}, 034323 (2021).

\bibitem{Caplan(2017)}
M.E.~Caplan and C.J.~Horowitz, Colloquium: Astromaterial science and nuclear pasta,
Rev. Mod. Phys. \textbf{89}, 041002 (2017).

\bibitem{Chamel(2017)}
N.~Chamel, Entrainment in Superfluid Neutron-Star Crusts: Hydrodynamic Description and Microscopic Origin,
J. Low Temp. Phys. \textbf{189}, 328 (2017).
\bibitem{Kashiwaba(2019)}
Y.~Kashiwaba and T.~Nakatsukasa, Self-consistent band calculation of the slab phase in the neutron-star crust,
Phys. Rev. C \textbf{100}, 035804 (2019).
\bibitem{Sekizawa(2022)}
K.~Sekizawa, S.~Kobayashi, and M.~Matsuo,
Time-dependent extension of the self-consistent band theory for neutron star matter: Anti-entrainment effects in the slab phase,
Phys. Rev. C \textbf{105}, 045807 (2022).

\end{thebibliography}
\end{document}